\documentclass[aps,prc,twocolumn,showpacs,preprintnumbers,amsmath,amssymb]{revtex4}

\usepackage{graphicx}
\usepackage{dcolumn}
\usepackage{bm}
\usepackage{color}
\usepackage{float}

\begin{document}

\title{Systematics of direct-$\alpha$ production with weakly and strongly bound projectiles}
\author{V. V. Parkar$^{1}$\footnote{vparkar@barc.gov.in}}
\author{V. Jha$^{1,2}$\footnote{vjha@barc.gov.in}}
\author{S. Kailas$^{1,3,4}$}
\affiliation{$^{1}$Nuclear Physics Division, Bhabha Atomic Research Centre, Mumbai-400085, India}
\affiliation{$^{2}$Homi Bhabha National Institute, Anushaktinagar, Mumbai-400094, India}
\affiliation{$^{3}$UM-DAE Centre for Excellence in Basic Sciences, Mumbai-400098, India}
\affiliation{$^{4}$Manipal Centre for Natural Sciences, Manipal Academy of Higher Education, Manipal-576104, India}

\begin{abstract}
The production of $\alpha$-particles in reactions using both the strongly and weakly bound projectiles at energies around the Coulomb barrier show several interesting features. To understand these, the role of various reaction mechanisms responsible for $\alpha$-production, such as non-capture breakup, capture of only one of the fragments subsequent to projectile breakup and their contribution to reaction cross sections have been investigated.  A systematic study of the $\alpha$-particle production based on available data for various projectile target systems have been performed and a classification based on projectile type is obtained.
\end{abstract}
\maketitle

\section{\label{sec:Intro} Introduction}
Enhanced production of $\alpha$-particles is a fascinating feature of the heavy-ion nuclear reactions at energies near or above the Coulomb barrier \cite{Britt61, Queb74, Casta1980}. It has been observed in experimental studies of reactions involving both the strongly bound (SBP) and weakly bound (WBP) projectiles. This phenomena has been utilized extensively for studies of clustering in light nuclei and investigations into various reaction mechanisms, such as breakup, transfer and its associated processes. For WBPs, especially for those having $\alpha$ + $x$ cluster structure, copious emission of $\alpha$-particles with cross sections, sometimes as large as the reaction cross section has been observed \cite{Santra12, Aguilera2000}. The weak binding and clustering effects that lead to an enhancement of breakup-transfer cross sections specially in reactions near the Coulomb barrier and its influence on the reactions dynamics is an important aspect of WBP induced reactions.  The contribution of breakup and transfer processes are expected to be magnified for the reactions involving unstable WBP namely, the radioactive ion beams (RIB) having lower $\alpha$ separation energies than stable WBP. Because of their extended radial distributions, investigations into $\alpha$-production in reactions with RIB's also offer possibilities to disentangle the effects due to binding energies and extended radial shapes. 

In recent years, several inclusive and exclusive measurements of $\alpha$-production have been performed that have focussed on understanding the relative contribution of different reaction processes. In general, $\alpha$ production cross section for WBP having $\alpha + x$ cluster structure can be written in terms of the following components 
\begin{equation}
\sigma_{\alpha} = \sigma_{\textrm{CF}} + \sigma_{\textrm{NCBU}} + \sigma_{\textrm{x-ICF/TR}} + \sigma_{\alpha_{\textrm{ICF/TR}}} + \sigma_{\textrm{QE(TR)}}
\end{equation} 
where, $\sigma_{\textrm{x-ICF/TR}}$, $\sigma_{\alpha_{\textrm{ICF/TR}}}$ correspond to cross sections of capture of $x$ fragment respectively or transfer of these fragments to the target. The components that originate from various transfer-breakup processes, such as, the capture of one of the fragments after the breakup leading to breakup-fusion or transfer of nucleon(s) to the continuum states of target provide the most dominant contributions in $\alpha$-production. In contrast, the contribution of other components namely, the non-capture breakup  (NCBU) and the nucleon or cluster transfer leading to low lying discrete states having quasielastic nature ($\sigma_{\textrm{QE(TR)}}$) are smaller in magnitude. Using the exclusive measurements by detecting both the primary breakup fragments, one can attempt to disentangle the contributions due to all these components. On the theoretical side, the NCBU can be effectively modelled using the continuum-discretized coupled channels (CDCC) calculations. In recent times absorption based models have been utilized to calculate the fragment-capture components which form the dominant part of the inclusive $\alpha$-production \cite{VVP16, VVP18, VVP18b}.   

The $\alpha$-production due to direct reaction mechanisms is found to be far more dominant compared to its production through the compound processes for the WBP's. The cross sections due to the complete fusion (CF) of projectile with the target ($\sigma_{\textrm{CF}}$) also contributes, however it is not significant especially in reactions for medium or heavy mass target nucleus. The CF process may also include the sequential capture of both the fragments following the breakup of projectile in two fragments. The difference of reaction and CF cross sections can be utilized for studying the importance of all direct processes contributing to the reaction cross section. In the present article, we perform a systematic study of non-compound $\alpha$-particle production with projectiles classified as SBP, WBP and RIB. We explain the observed universal behaviour of the non-compound $\alpha$-production cross sections observed at energies near the Coulomb barrier with $^{6,7}$Li projectiles \cite{Pfei73, Pako03, Kumawat10, Santra12, Pandit17}. We also discuss about the reaction channels responsible for the observed enhanced $\alpha$ production in these cases.
\section*{Systematics of $\alpha$-particle production for reactions with strongly bound projectiles}
\begin{figure}
\begin{center}
\vspace{-0.4cm}
\includegraphics[trim = 0.3cm 17.8cm 11.0cm 1.5cm, clip, width = 85mm]{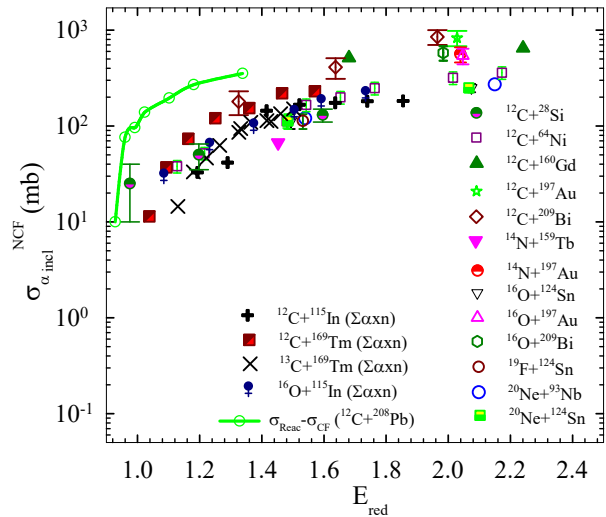}
\caption{\label{alphaprod_SBP} Systematical behaviour of  inclusive $\alpha$ production cross sections due to non-CF processes in reactions with SBP systems as a function of reduced energy. The plot also includes the data for residue measurements using $\Sigma \alpha xn$ channels. The line shows $\sigma_{Reac}$-$\sigma_{CF}$ data for $^{12}$C+$^{208}$Pb system.}
\end{center}
\end{figure}

Large data sets exist for the the inclusive $\alpha$ measurements for SBP systems with various targets. Inclusive $\alpha$ production data have been measured for reactions using SBP's $^{12,13}$C, $^{14}$N, $^{16}$O, $^{19}$F and $^{20}$Ne with many targets \cite{Britt61, Siwek79n, Hugi81, Tricoire82, Balster87a,  Parker87, Li88}. It is useful to first separate out the yield of evaporation $\alpha$ particles due to the CF contribution. The CF part has been estimated from the statistical model calculations using code PACE2 \cite{Gavr80} and non-CF inclusive $\alpha$ production cross sections (${\sigma _{\alpha_{incl.}}^{NCF}}$) have been determined.  Since, we are comparing data with different projectile-target systems in the vicinity of the Coulomb barrier(CB), the $\rm {c.m.}$ energies are reduced as $E_{\rm red}$ as defined below, 
\begin{equation}
 E_{\rm red} = E_{\rm c.m.}/[(Z_{\rm P}Z_{\rm T})/(A_{\rm P}^{1/3} + A_{\rm T} ^{1/3})]
\end{equation}
The plot of ${\sigma _{\alpha_{incl}}^{NCF}}$ for the SBP's  with reduced energy $E_{red}$ for various SBP systems is shown in Fig.\ \ref{alphaprod_SBP}. The data for residue measurements of $\Sigma \alpha xn$ channels associated with emission of one or more $\alpha$ particles is also included. An increase in ${\sigma _{\alpha_{incl}}^{NCF}}$ with incident energy and a reasonable similarity in the behaviour for different systems is observed. A comparison of ${\sigma _{\alpha_{incl}}^{NCF}}$ for SBP is made with the measured data of $\sigma_{Reac}$-$\sigma_{CF}$ for $^{12}$C+$^{208}$Pb system \cite{Santra01, Mukherjee07}.  It has been shown that the quantity $\sigma_{Reac}$-$\sigma_{CF}$ has a resonable systematic dependence for all SBP's. This quantity is found to be much larger than ${\sigma _{\alpha_{incl}}^{NCF}}$, suggesting that other processes such as, inelastic and transfer or incomplete fusion (ICF) processes due to non-$\alpha$-emitting channels may also contribute significantly to the reaction cross section for SBP's.  
\section*{Systematics of $\alpha$-particle production for reactions with stable weakly bound projectiles}
For the case of WBP's, relatively larger inclusive $\alpha$ cross section have been measured specially, in reactions using the $^{6,7}$Li projectiles on several targets \cite{Pfei73, Hugi81, Pako03, Souz09, Shri06, Kumawat10, Chatto16, Prad13, Ost72, Sign01, Santra12, Pako05, Pandit17}. In general, $\alpha$ fragment arising from the projectile breakup interacts relatively less when compared to the other fragment, leading to partial-capture and observation of a large $\alpha$ emission; \textit{viz}, in the case of $^7$Li, triton fusion is more favoured compared to $\alpha$ fusion, which leads to dominant emission of $\alpha$ particles. 

For $^6$Li induced reactions around the CB, inclusive $\alpha$ production cross sections have been found to be very dominant at sub-barrier energies. The contribution of pure compound processes leading to CF estimated by the statistical model calculations is much less compared to the direct processes as shown for $^6$Li + $^{90}$Zr system \cite{Kumawat10}. The dominant contribution to ${\sigma _{\alpha_{incl}}^{NCF}}$ is given by $d$-capture and/or $d$ cluster transfer \cite{Santra12, Souza2010} however,  1$\textit{n}$ stripping \cite{Pradhan11} channel also contributes to $\alpha$ production. For $^7$Li projectile systems, capture of $t$ cluster or direct $t$-transfer have been observed to contribute dominantly to $\alpha$ production. NCBU cross sections including transfer followed by breakup such as, 1$n$ stripping \cite{Shri06, Pand16, Chattopadhyay2018} and 1$p$ pickup \cite{Pand16, Chatto18b} found to contribute around $\approx$10 \% to ${\sigma _{\alpha_{incl}}^{NCF}}$ for both $^6$Li and $^7$Li projectiles with medium and heavy mass targets. 
\begin{figure*}
\begin{center}
\vspace{-0.4cm}
\includegraphics[trim = 0.4cm 7.7cm 3.4cm 8.0cm, clip, width = 170mm]{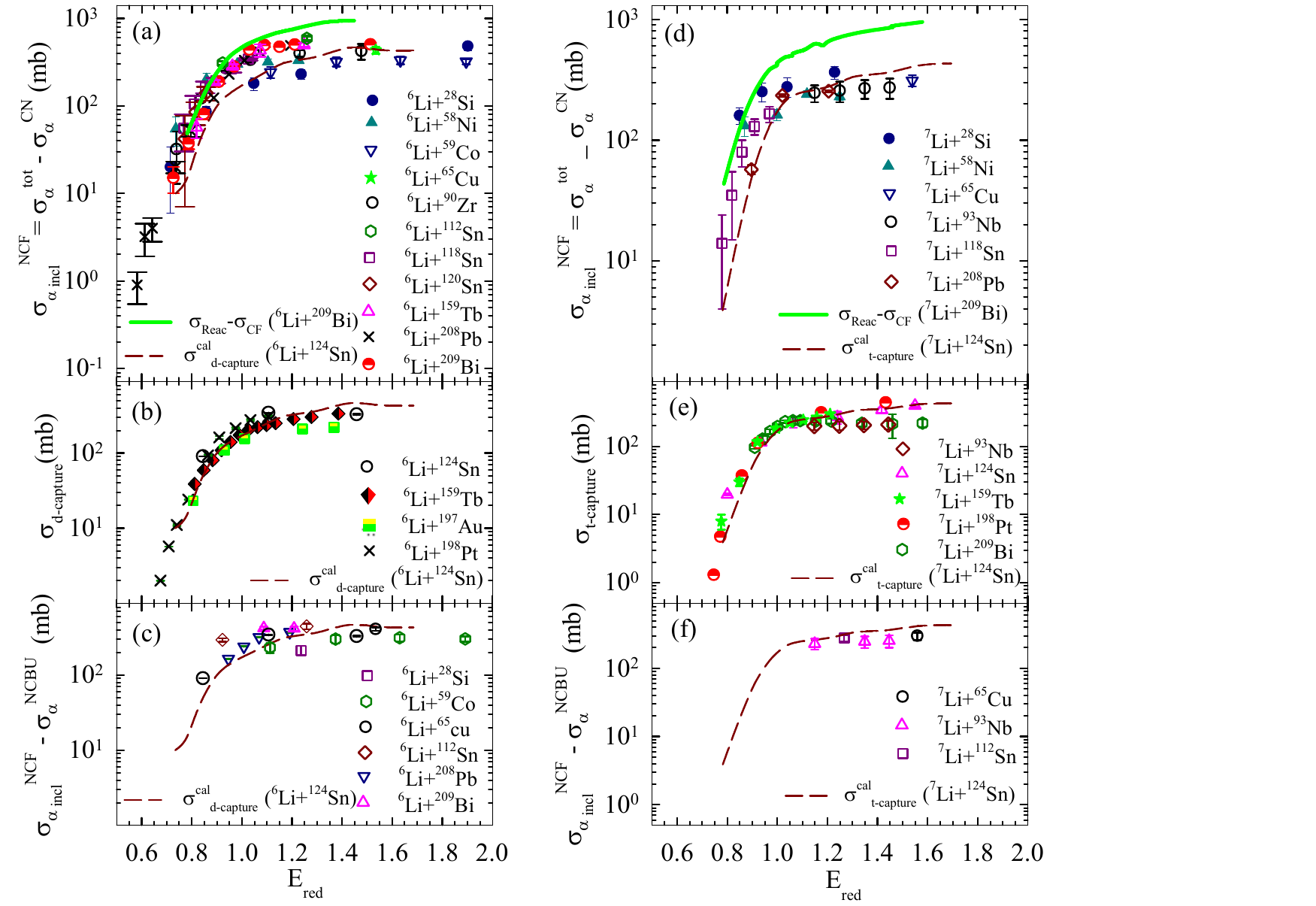}
\caption{\label{alphaprod_67Li} (a,d) Systematical behaviour of inclusive $\alpha$ production cross sections due to non-CF processes, (b,e) Measured $d$ ($t$)-cature cross sections, (c,f) Difference between measured ${\sigma _{\alpha_{incl}}^{NCF}}$ and ${\sigma _{\alpha}^{NCBU}}$ with $^6$Li ($^7$Li) projectile as a function of reduced energy. The dashed lines in (a-c) and (d-f) are results of d-capture and t-capture calculations performed for $^6$Li+$^{124}$Sn and $^7$Li+$^{124}$Sn systems, respectively. The solid lines in (a) and (d) are calculated values of $\sigma_{Reac}$-$\sigma_{CF}$ for $^6$Li+$^{209}$Bi and $^7$Li+$^{209}$Bi systems, respectively.}
\end{center}
\end{figure*}

We have made comparative studies of non-CF $\alpha$ production with $^{6,7}$Li projectiles. As can be seen from Fig.\ \ref{alphaprod_67Li} (a), universal behaviour is found in ${\sigma _{\alpha_{incl}}^{NCF}}$ for medium to heavy target nuclei. For the light target nuclei, there is a larger CF contribution which leads to the lower values of ${\sigma _{\alpha_{incl}}^{NCF}}$ \cite{Pako03}. The difference of the reaction and CF cross sections have been found to be nearly the same for different systems as a function of E$_{red}$ for values at energies approximately twice the CB for $^6$Li induced reactions \cite{Kumawat10, Santra12}. From Fig.\ \ref{alphaprod_67Li}(a), it is seen that the calculated values of $\sigma_{Reac}$-$\sigma_{CF}$ for $^{6}$Li+$^{209}$Bi system shown by solid line matches well with measured data of ${\sigma _{\alpha_{incl}}^{NCF}}$ for all $^6$Li target systems. Here, $\sigma_{Reac}$ is calculated using Sau-Paulo potential \cite{Chamon02} and $\sigma_{CF}$ is calculated CF cross section taken from Ref.\ \cite{VVP16} for $^{6}$Li+$^{209}$Bi system. It can be concluded that non-CF $\alpha$ production is the only important process apart from CF in $^{6}$Li induced reactions. 

We have also shown the calculated $d$-capture cross sections for $^{6}$Li+$^{124}$Sn system from Ref.\ \cite{VVP18b} in Fig.\ \ref{alphaprod_67Li} (a) and the data for it on various targets \cite{VVP18b, Pradhan11, Pals14, Shrivastava09} in Fig.\ \ref{alphaprod_67Li} (b). The calculations of d-capture cross sections are performed using CDCC-absorption model as described in Ref.\ \cite{VVP18b} and a good description of measured d-capture data is obtained. It is found that the $d$-capture data and calculations for it underpredict the ${\sigma _{\alpha_{incl}}^{NCF}}$, suggesting there are mechanisms other than $d$-capture that contribute to the direct $\alpha$ production, specially for targets with A$>$90. In particular, the $n$ transfer from $^6$Li could also contribute to $\alpha$ production as shown in Ref.\ \cite{Pradhan11}. 

Similar to $^6$Li induced reactions, $^7$Li induced reactions also show universal behaviour in ${\sigma _{\alpha_{incl}}^{NCF}}$ as shown in Fig.\ \ref{alphaprod_67Li}(d). It is also seen that the $\alpha$ production is more with $^6$Li than with $^7$Li projectile due to lower breakup threshold of $^6$Li. However, the calculated values of $\sigma_{Reac}$-$\sigma_{CF}$ for $^{7}$Li+$^{209}$Bi system is found to be much larger than ${\sigma _{\alpha_{incl}}^{NCF}}$, which suggests that inelastic and other transfer processes may also contribute significantly in reaction cross section for $^7$Li case. As before, the $\sigma_{Reac}$ is calculated using Sau-Paulo potential \cite{Chamon02} and $\sigma_{CF}$ is calculated CF cross section taken from Ref.\ \cite{VVP16} for $^{7}$Li+$^{209}$Bi system. 

The CDCC-absorption model calculations have been shown to provide a good description of $t$-capture cross sections as described in Ref.\ \cite{VVP18b} for $^{7}$Li+$^{124}$Sn system. In Fig.\ \ref{alphaprod_67Li}(e) we have compared these $t$-capture calculations with $t$ capture data on various targets \cite{VVP18, Pandit17, Brod75, Ara13, Dasgupta04}. The $t$ capture data match well with the ${\sigma _{\alpha_{incl}}^{NCF}}$ showing that non-compound $\alpha$-production dominantly orginates from this path. 

An indirect way to estimate the $d$-capture and $t$-capture cross sections for $^6$Li and $^7$Li systems is the subtraction of NCBU cross sections (${\sigma _{\alpha}^{NCBU}}$) from the ${\sigma _{\alpha_{incl}}^{NCF}}$ measured for various targets \cite{Pako06, Souz09, Shri06, Chatto16, Sign03, Sant09, Pand16, Chattopadhyay2018, Chatto18b}. As shown in Fig.\ \ref{alphaprod_67Li}(c) and Fig.\ \ref{alphaprod_67Li}(f), $d$-capture and $t$-capture cross sections determined this way can be well explained by the $d$- and $t$-capture calculations and this provides a useful way to determine the $d$-ICF and $t$-ICF cross sections for these systems.

\begin{figure}
\begin{center}
\includegraphics[trim = 0.6cm 20.3cm 14.0cm 1.0cm, clip, width = 85mm]{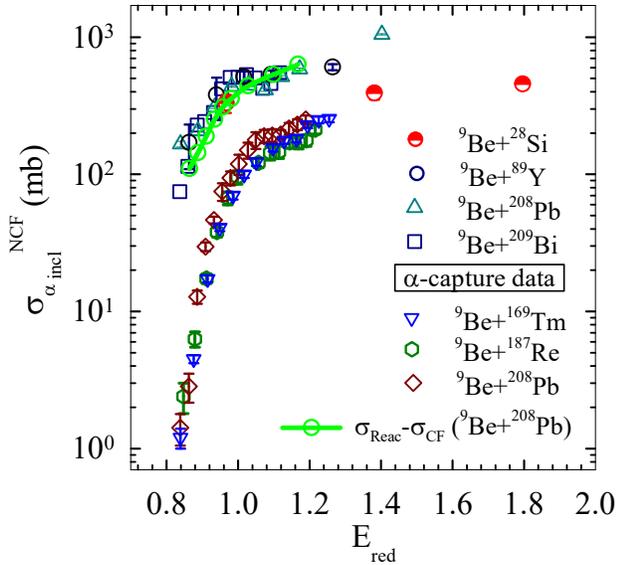}
\caption{\label{alphaprod_9Be} Systematical behaviour of inclusive $\alpha$ production cross sections due to non-CF processes in reactions with $^9$Be projectile as a function of reduced energy is shown along with $\alpha$-capture cross sections. The line shows $\sigma_{Reac}$-$\sigma_{CF}$ data for $^{9}$Be+$^{208}$Pb system.}
\end{center}
\end{figure}
For the $^9$Be projectile, the $\alpha$ production may correspond to $\alpha$-capture and $n$-capture leading to production of one and two $\alpha$ particles in a single event. The $\alpha$ production in this case may be given as 
\begin{equation}
\sigma_{\alpha} = \sigma_{{\alpha}-capture} + 2 (\sigma_{n-capture} + \sigma _{n-TR})
\end{equation}  
where,  $\sigma _{n-TR}$ is the contribution due to neutron-transfer. We have extracted the non-CF $\alpha$-production cross sections from the available inclusive $\alpha$-production data for $^9$Be projectile on various target systems \cite{Hugi81, Sig04, Wooll03, Pals14b} and shown in Fig.\ \ref{alphaprod_9Be}. A systematic behaviour of ${\sigma _{\alpha_{incl}}^{NCF}}$ for all target systems, similar to those observed for $^{6,7}$Li projectiles is seen. This data can be well described by measured $\sigma_{Reac}$-$\sigma_{CF}$ values for $^9$Be + $^{208}$Pb system \cite{Dasgupta04, Yu10}. The non-CF $\alpha$ production in this case is much larger compared to the measured $\alpha$-ICF for various targets \cite{Fang15, Dasgupta04}. This is due to the dominant contribution of other modes of $\alpha$ production including the $n$-transfer which contributes signicantly. While the $1n$ transfer to the low lying states of $^{209}$Pb populated in $^9$Be + $^{208}$Pb reaction was found to be smaller, the $1n$ transfer to the high-lying states may give a dominant contribution to the inclusive $\alpha$ production in this case \cite{Jha14}.  
\section*{Systematics of $\alpha$-particle production for reactions with RIB's}
The $\alpha$ production in reactions with the unstable WBP namely the RIB's, have been also measured for several targets at energies around the CB. The RIB's have binding energies that range from a few MeV to a very low value of only a few hundred keV. In addition, RIB's are also characterized by extended radial distributions including some of them having the halo structure. A large cross section of the production of $\alpha$ particles is observed in reactions with $^6$He projectile on several targets \cite{Piet04, Scu11, Standy13, Aguilera2000, Kol02}. A large inclusive $\alpha$ yield found at near-barrier energies \cite{Aguilera2000, Aguilera2001} can be ascribed to the weak binding of the halo neutrons that favors the dissociation of the $^6$He projectile in the nuclear and Coulomb field of the target. The inclusive $\alpha$ cross sections are much larger than the fusion cross section at these energies. The measured energy spectra and the angular distribution of the inclusive $\alpha$ production channels for the $^6$He + $^{208}$Pb system at energies around the CB have been explained in terms of the coupled channels calculations using transfer to the continuum method \cite{ESCRIG2007} suggesting 2n-transfer plays a key role in $\alpha$ production. 

\begin{figure}
\begin{center}
\includegraphics[trim = 0.8cm 17.7cm 9.9cm 1.6cm, clip, width = 85mm]{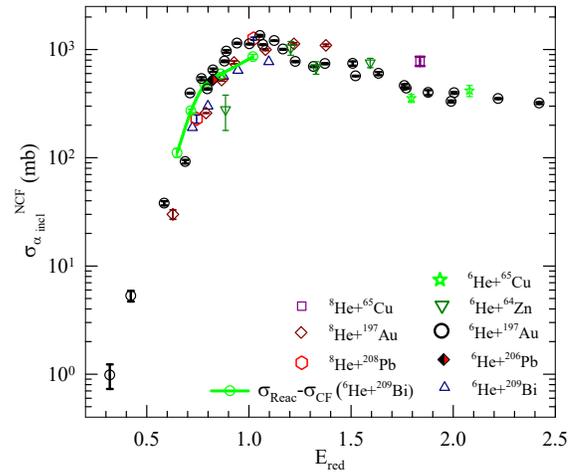}
\caption{\label{alphaprod_RIB} Systematical behaviour of inclusive $\alpha$ production cross sections due to non-CF processes in reactions with $^{6,8}$He projectiles as a function of reduced energy. The line shows $\sigma_{Reac}$-$\sigma_{CF}$ data for $^{6}$He+$^{209}$Bi system.}
\end{center}
\end{figure}

A universal behaviour similar to that obtained for stable WBP for the non-CF part of $\alpha$ production cross sections for systems with $^6$He and $^8$He projectiles is shown in Fig.\ \ref{alphaprod_RIB}. 
In this figure, exclusive data of the n-transfer for $^{6}$He + $^{65}$Cu \cite{Navi04, Chat08}, $^{197}$Au \cite{Peni07}, and $^{8}$He + $^{65}$Cu \cite{Lema11}, $^{197}$Au \cite{Lema09} systems is included while the remaining are from inclusive $\alpha$ measurements \cite{Piet04, Scu11, Standy13, Aguilera2000, Kol02, Marq18}.  However, there is a decrease in ${\sigma _{\alpha_{incl}}^{NCF}}$ at energies above the barrier in contrast to stable WBP and SBP systems. The  experimental data of ${\sigma _{\alpha_{incl}}^{NCF}}$ can be well described by the measured $\sigma_{Reac}$-$\sigma_{CF}$ data for $^{6}$He+$^{209}$Bi system \cite{Kol02} at energies around the CB as shown in Fig.\ \ref{alphaprod_RIB} similar as $^{6}$Li case suggesting the negligible contribution of any other channels. 
\section*{Comparison of $\alpha$-particle production in SBP, WBP and RIB}
\begin{figure}
\begin{center}
\includegraphics[trim = 2.3cm 19.3cm 8.5cm 1.1cm, clip, width = 85mm]{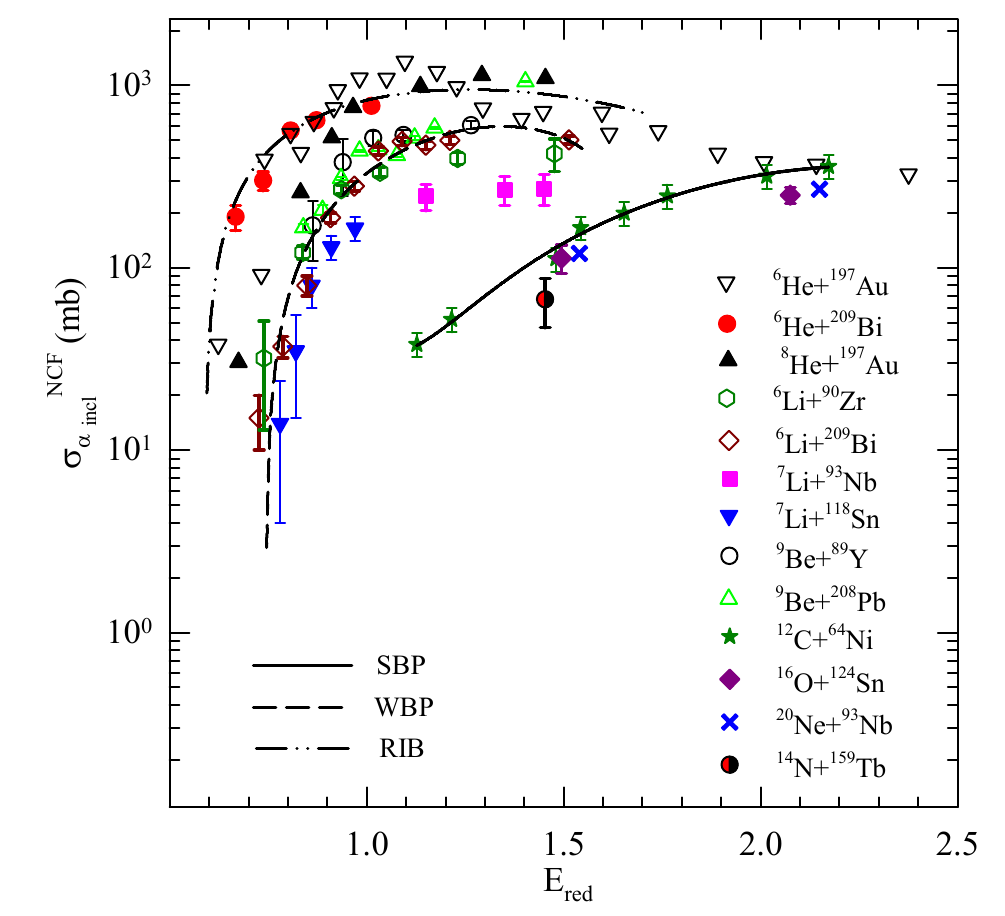}
\caption{\label{sbpwbp_alpha} Systematic comparison of inclusive $\alpha$ production cross sections due to non-CF processes for different nuclear systems in three categories: (i) SBP, (ii) stable WBP, and (iii) RIB. Lines are guide to an eye.}
\end{center}
\end{figure}

Next, we perform a  comparative study  of ${\sigma _{\alpha_{incl}}^{NCF}}$ for all three types of projectile systems categorized as, (i) SBP, (ii) stable WBP, and (iii) RIB is shown in Fig.\ \ref{sbpwbp_alpha}. There is a characteristic difference observed in ${\sigma _{\alpha_{incl}}^{NCF}}$ for these projectile systems. A similar behaviour was observed for the reaction cross sections \cite{Kola16}, where larger values are seen for RIB compared to the values for stable WBP, which are in turn larger than the values for SBP. It can be seen that the energy values where the ${\sigma _{\alpha_{incl}}^{NCF}}$ saturate are much higher for SBP ($\approx$ 2V$_B$) than the value for stable WBP and RIB ($\approx$ 1.2V$_B$). It can be concluded that for the RIB's, the  smaller binding energies coupled with extended radial shapes contribute to larger values of both $\sigma_{Reac}$ and ${\sigma _{\alpha_{incl}}^{NCF}}$. Somewhat similar beahviour in inclusive $\alpha$ cross sections was reported earlier with weakly bound light projectiles $^{6,7}$Li, $^9$Be and $^6$He using heavy $^{208}$Pb and $^{209}$Bi targets \cite{Kol01}. 
\section*{Summary}
In summary, we have investigated the systematics of non-CF $\alpha$-particle production for various projectile systems having predominantly $\alpha$ + $x$ cluster structure. The non-CF $\alpha$ particle production alone along with CF is not able to completely explain the reaction cross sections for the SBP systems. In contrast, for the $^6$Li WBP system cases, the reaction cross sections are completely explained by sum of CF cross sections and cross sections for non-CF $\alpha$ production. The non-CF $\alpha$ production in $^{6}$Li is mainly due to $d$-capture (transfer), however other processes such as 1$n$ stripping also contribute in the $\alpha$-particle production. In contrast for the $^7$Li case, only the $t$-capture (transfer) is sufficient to explain $\alpha$ production, but other channels also contribute significantly to the reaction cross section. Quantitative description of $d$ and $t$-capture  for $^6$Li and $^7$Li projectile systems respectively, can be obtained by CDCC based absorption model calculations. A comparative study among the SBP, WBP and RIB projectile systems show that the $\alpha$ production is more with RIB than stable WBP cases which in turn is higher than the SBP case.

\begin{thebibliography}{62}
\expandafter\ifx\csname natexlab\endcsname\relax\def\natexlab#1{#1}\fi
\expandafter\ifx\csname bibnamefont\endcsname\relax
  \def\bibnamefont#1{#1}\fi
\expandafter\ifx\csname bibfnamefont\endcsname\relax
  \def\bibfnamefont#1{#1}\fi
\expandafter\ifx\csname citenamefont\endcsname\relax
  \def\citenamefont#1{#1}\fi
\expandafter\ifx\csname url\endcsname\relax
  \def\url#1{\texttt{#1}}\fi
\expandafter\ifx\csname urlprefix\endcsname\relax\def\urlprefix{URL }\fi
\providecommand{\bibinfo}[2]{#2}
\providecommand{\eprint}[2][]{\url{#2}}

\bibitem[{\citenamefont{{Harold C. Britt} and {Arthur R.
  Quinton}}(1961)}]{Britt61}
\bibinfo{author}{\bibnamefont{{Harold C. Britt}}} \bibnamefont{and}
  \bibinfo{author}{\bibnamefont{{Arthur R. Quinton}}}, \bibinfo{journal}{Phys.
  Rev.} \textbf{\bibinfo{volume}{124}}, \bibinfo{pages}{877}
  (\bibinfo{year}{1961}).

\bibitem[{\citenamefont{Qu\'ebert et~al.}(1974)\citenamefont{Qu\'ebert, Frois,
  Marquez, Sousbie, Ost, Bethge, and Gruber}}]{Queb74}
\bibinfo{author}{\bibfnamefont{J.~L.} \bibnamefont{Qu\'ebert}},
  \bibinfo{author}{\bibfnamefont{B.}~\bibnamefont{Frois}},
  \bibinfo{author}{\bibfnamefont{L.}~\bibnamefont{Marquez}},
  \bibinfo{author}{\bibfnamefont{G.}~\bibnamefont{Sousbie}},
  \bibinfo{author}{\bibfnamefont{R.}~\bibnamefont{Ost}},
  \bibinfo{author}{\bibfnamefont{K.}~\bibnamefont{Bethge}}, \bibnamefont{and}
  \bibinfo{author}{\bibfnamefont{G.}~\bibnamefont{Gruber}},
  \bibinfo{journal}{Phys. Rev. Lett.} \textbf{\bibinfo{volume}{32}},
  \bibinfo{pages}{1136} (\bibinfo{year}{1974}).

\bibitem[{\citenamefont{Castaneda et~al.}(1980)\citenamefont{Castaneda, Smith,
  Singh, and Karwowski}}]{Casta1980}
\bibinfo{author}{\bibfnamefont{C.~M.} \bibnamefont{Castaneda}},
  \bibinfo{author}{\bibfnamefont{H.~A.} \bibnamefont{Smith}},
  \bibinfo{author}{\bibfnamefont{P.~P.} \bibnamefont{Singh}}, \bibnamefont{and}
  \bibinfo{author}{\bibfnamefont{H.}~\bibnamefont{Karwowski}},
  \bibinfo{journal}{Phys. Rev. C} \textbf{\bibinfo{volume}{21}},
  \bibinfo{pages}{179} (\bibinfo{year}{1980}).

\bibitem[{\citenamefont{Santra et~al.}(2012)\citenamefont{Santra, Kailas,
  Parkar, Ramachandran, Jha, Chatterjee, Rath, and Parihari}}]{Santra12}
\bibinfo{author}{\bibfnamefont{S.}~\bibnamefont{Santra}},
  \bibinfo{author}{\bibfnamefont{S.}~\bibnamefont{Kailas}},
  \bibinfo{author}{\bibfnamefont{V.~V.} \bibnamefont{Parkar}},
  \bibinfo{author}{\bibfnamefont{K.}~\bibnamefont{Ramachandran}},
  \bibinfo{author}{\bibfnamefont{V.}~\bibnamefont{Jha}},
  \bibinfo{author}{\bibfnamefont{A.}~\bibnamefont{Chatterjee}},
  \bibinfo{author}{\bibfnamefont{P.~K.} \bibnamefont{Rath}}, \bibnamefont{and}
  \bibinfo{author}{\bibfnamefont{A.}~\bibnamefont{Parihari}},
  \bibinfo{journal}{Phys. Rev.} \textbf{\bibinfo{volume}{C 85}},
  \bibinfo{pages}{014612} (\bibinfo{year}{2012}).

\bibitem[{\citenamefont{Aguilera et~al.}(2000)\citenamefont{Aguilera, Kolata,
  Nunes, Becchetti, DeYoung, Goupell, Guimar\~aes, Hughey, Lee, Lizcano
  et~al.}}]{Aguilera2000}
\bibinfo{author}{\bibfnamefont{E.~F.} \bibnamefont{Aguilera}},
  \bibinfo{author}{\bibfnamefont{J.~J.} \bibnamefont{Kolata}},
  \bibinfo{author}{\bibfnamefont{F.~M.} \bibnamefont{Nunes}},
  \bibinfo{author}{\bibfnamefont{F.~D.} \bibnamefont{Becchetti}},
  \bibinfo{author}{\bibfnamefont{P.~A.} \bibnamefont{DeYoung}},
  \bibinfo{author}{\bibfnamefont{M.}~\bibnamefont{Goupell}},
  \bibinfo{author}{\bibfnamefont{V.}~\bibnamefont{Guimar\~aes}},
  \bibinfo{author}{\bibfnamefont{B.}~\bibnamefont{Hughey}},
  \bibinfo{author}{\bibfnamefont{M.~Y.} \bibnamefont{Lee}},
  \bibinfo{author}{\bibfnamefont{D.}~\bibnamefont{Lizcano}},
  \bibnamefont{et~al.}, \bibinfo{journal}{Phys. Rev. Lett.}
  \textbf{\bibinfo{volume}{84}}, \bibinfo{pages}{5058} (\bibinfo{year}{2000}).

\bibitem[{\citenamefont{Parkar et~al.}(2016)\citenamefont{Parkar, Jha, and
  Kailas}}]{VVP16}
\bibinfo{author}{\bibfnamefont{V.~V.} \bibnamefont{Parkar}},
  \bibinfo{author}{\bibfnamefont{V.}~\bibnamefont{Jha}}, \bibnamefont{and}
  \bibinfo{author}{\bibfnamefont{S.}~\bibnamefont{Kailas}},
  \bibinfo{journal}{Phys. Rev. C} \textbf{\bibinfo{volume}{94}},
  \bibinfo{pages}{024609} (\bibinfo{year}{2016}).

\bibitem[{\citenamefont{Parkar et~al.}(2018{\natexlab{a}})\citenamefont{Parkar,
  {Sushil K. Sharma}, Palit, Upadhyaya, Shrivastava, Pandit, Mahata, Jha,
  Santra, Ramachandran et~al.}}]{VVP18}
\bibinfo{author}{\bibfnamefont{V.~V.} \bibnamefont{Parkar}},
  \bibinfo{author}{\bibnamefont{{Sushil K. Sharma}}},
  \bibinfo{author}{\bibfnamefont{R.}~\bibnamefont{Palit}},
  \bibinfo{author}{\bibfnamefont{S.}~\bibnamefont{Upadhyaya}},
  \bibinfo{author}{\bibfnamefont{A.}~\bibnamefont{Shrivastava}},
  \bibinfo{author}{\bibfnamefont{S.~K.} \bibnamefont{Pandit}},
  \bibinfo{author}{\bibfnamefont{K.}~\bibnamefont{Mahata}},
  \bibinfo{author}{\bibfnamefont{V.}~\bibnamefont{Jha}},
  \bibinfo{author}{\bibfnamefont{S.}~\bibnamefont{Santra}},
  \bibinfo{author}{\bibfnamefont{K.}~\bibnamefont{Ramachandran}},
  \bibnamefont{et~al.}, \bibinfo{journal}{Phys. Rev. C}
  \textbf{\bibinfo{volume}{97}}, \bibinfo{pages}{014607}
  (\bibinfo{year}{2018}{\natexlab{a}}).

\bibitem[{\citenamefont{Parkar et~al.}(2018{\natexlab{b}})\citenamefont{Parkar,
  Pandit, Shrivastava, Palit, Mahata, Jha, Ramachandran, {Shilpi Gupta},
  Santra, {Sushil K. Sharma} et~al.}}]{VVP18b}
\bibinfo{author}{\bibfnamefont{V.~V.} \bibnamefont{Parkar}},
  \bibinfo{author}{\bibfnamefont{S.~K.} \bibnamefont{Pandit}},
  \bibinfo{author}{\bibfnamefont{A.}~\bibnamefont{Shrivastava}},
  \bibinfo{author}{\bibfnamefont{R.}~\bibnamefont{Palit}},
  \bibinfo{author}{\bibfnamefont{K.}~\bibnamefont{Mahata}},
  \bibinfo{author}{\bibfnamefont{V.}~\bibnamefont{Jha}},
  \bibinfo{author}{\bibfnamefont{K.}~\bibnamefont{Ramachandran}},
  \bibinfo{author}{\bibnamefont{{Shilpi Gupta}}},
  \bibinfo{author}{\bibfnamefont{S.}~\bibnamefont{Santra}},
  \bibinfo{author}{\bibnamefont{{Sushil K. Sharma}}}, \bibnamefont{et~al.},
  \bibinfo{journal}{Phys. Rev. C} \textbf{\bibinfo{volume}{98}},
  \bibinfo{pages}{014601} (\bibinfo{year}{2018}{\natexlab{b}}).

\bibitem[{\citenamefont{Pfeiffer et~al.}(1973)\citenamefont{Pfeiffer, Speth,
  and Bethge}}]{Pfei73}
\bibinfo{author}{\bibfnamefont{K.~O.} \bibnamefont{Pfeiffer}},
  \bibinfo{author}{\bibfnamefont{E.}~\bibnamefont{Speth}}, \bibnamefont{and}
  \bibinfo{author}{\bibfnamefont{K.}~\bibnamefont{Bethge}},
  \bibinfo{journal}{Nucl. Phys. A} \textbf{\bibinfo{volume}{206}},
  \bibinfo{pages}{545} (\bibinfo{year}{1973}).

\bibitem[{\citenamefont{Pakou et~al.}(2003)\citenamefont{Pakou, Alamanos,
  Gillibert, Kokkoris, Kossionides, Lagoyannis, Nicolis, Papachristodoulou,
  Patiris, Pierroutsakou et~al.}}]{Pako03}
\bibinfo{author}{\bibfnamefont{A.}~\bibnamefont{Pakou}},
  \bibinfo{author}{\bibfnamefont{N.}~\bibnamefont{Alamanos}},
  \bibinfo{author}{\bibfnamefont{A.}~\bibnamefont{Gillibert}},
  \bibinfo{author}{\bibfnamefont{M.}~\bibnamefont{Kokkoris}},
  \bibinfo{author}{\bibfnamefont{S.}~\bibnamefont{Kossionides}},
  \bibinfo{author}{\bibfnamefont{A.}~\bibnamefont{Lagoyannis}},
  \bibinfo{author}{\bibfnamefont{N.~G.} \bibnamefont{Nicolis}},
  \bibinfo{author}{\bibfnamefont{C.}~\bibnamefont{Papachristodoulou}},
  \bibinfo{author}{\bibfnamefont{D.}~\bibnamefont{Patiris}},
  \bibinfo{author}{\bibfnamefont{D.}~\bibnamefont{Pierroutsakou}},
  \bibnamefont{et~al.}, \bibinfo{journal}{Phys. Rev. Lett.}
  \textbf{\bibinfo{volume}{90}}, \bibinfo{pages}{202701}
  (\bibinfo{year}{2003}).

\bibitem[{\citenamefont{Kumawat et~al.}(2010)\citenamefont{Kumawat, Jha,
  Parkar, Roy, Santra, Kumar, Dutta, Shukla, Pant, Mohanty et~al.}}]{Kumawat10}
\bibinfo{author}{\bibfnamefont{H.}~\bibnamefont{Kumawat}},
  \bibinfo{author}{\bibfnamefont{V.}~\bibnamefont{Jha}},
  \bibinfo{author}{\bibfnamefont{V.~V.} \bibnamefont{Parkar}},
  \bibinfo{author}{\bibfnamefont{B.~J.} \bibnamefont{Roy}},
  \bibinfo{author}{\bibfnamefont{S.}~\bibnamefont{Santra}},
  \bibinfo{author}{\bibfnamefont{V.}~\bibnamefont{Kumar}},
  \bibinfo{author}{\bibfnamefont{D.}~\bibnamefont{Dutta}},
  \bibinfo{author}{\bibfnamefont{P.}~\bibnamefont{Shukla}},
  \bibinfo{author}{\bibfnamefont{L.~M.} \bibnamefont{Pant}},
  \bibinfo{author}{\bibfnamefont{A.~K.} \bibnamefont{Mohanty}},
  \bibnamefont{et~al.}, \bibinfo{journal}{Phys. Rev. C}
  \textbf{\bibinfo{volume}{81}}, \bibinfo{pages}{054601}
  (\bibinfo{year}{2010}).

\bibitem[{\citenamefont{Pandit et~al.}(2017)\citenamefont{Pandit, Shrivastava,
  Mahata, Parkar, Palit, Keeley, Rout, Kumar, Ramachandran, Bhattacharyya
  et~al.}}]{Pandit17}
\bibinfo{author}{\bibfnamefont{S.~K.} \bibnamefont{Pandit}},
  \bibinfo{author}{\bibfnamefont{A.}~\bibnamefont{Shrivastava}},
  \bibinfo{author}{\bibfnamefont{K.}~\bibnamefont{Mahata}},
  \bibinfo{author}{\bibfnamefont{V.~V.} \bibnamefont{Parkar}},
  \bibinfo{author}{\bibfnamefont{R.}~\bibnamefont{Palit}},
  \bibinfo{author}{\bibfnamefont{N.}~\bibnamefont{Keeley}},
  \bibinfo{author}{\bibfnamefont{P.~C.} \bibnamefont{Rout}},
  \bibinfo{author}{\bibfnamefont{A.}~\bibnamefont{Kumar}},
  \bibinfo{author}{\bibfnamefont{K.}~\bibnamefont{Ramachandran}},
  \bibinfo{author}{\bibfnamefont{S.}~\bibnamefont{Bhattacharyya}},
  \bibnamefont{et~al.}, \bibinfo{journal}{Phys. Rev. C}
  \textbf{\bibinfo{volume}{96}}, \bibinfo{pages}{044616}
  (\bibinfo{year}{2017}).

\bibitem[{\citenamefont{Siwek-Wilczynska
  et~al.}(1979)\citenamefont{Siwek-Wilczynska, {E. H. du Marchie van
  Voorthuysen}, van Popta, Siemssen, and Wilczynski}}]{Siwek79n}
\bibinfo{author}{\bibfnamefont{K.}~\bibnamefont{Siwek-Wilczynska}},
  \bibinfo{author}{\bibnamefont{{E. H. du Marchie van Voorthuysen}}},
  \bibinfo{author}{\bibfnamefont{J.}~\bibnamefont{van Popta}},
  \bibinfo{author}{\bibfnamefont{R.~H.} \bibnamefont{Siemssen}},
  \bibnamefont{and}
  \bibinfo{author}{\bibfnamefont{J.}~\bibnamefont{Wilczynski}},
  \bibinfo{journal}{Nucl. Phys. A} \textbf{\bibinfo{volume}{330}},
  \bibinfo{pages}{150} (\bibinfo{year}{1979}).

\bibitem[{\citenamefont{Hugi et~al.}(1981)\citenamefont{Hugi, Lang, M{\"
  u}ller, Ungricht, Bodek, Jarczyk, Kamys, Magiera, Strza\l{}kowski, and
  Willim}}]{Hugi81}
\bibinfo{author}{\bibfnamefont{M.}~\bibnamefont{Hugi}},
  \bibinfo{author}{\bibfnamefont{J.}~\bibnamefont{Lang}},
  \bibinfo{author}{\bibfnamefont{R.}~\bibnamefont{M{\" u}ller}},
  \bibinfo{author}{\bibfnamefont{E.}~\bibnamefont{Ungricht}},
  \bibinfo{author}{\bibfnamefont{K.}~\bibnamefont{Bodek}},
  \bibinfo{author}{\bibfnamefont{L.}~\bibnamefont{Jarczyk}},
  \bibinfo{author}{\bibfnamefont{B.}~\bibnamefont{Kamys}},
  \bibinfo{author}{\bibfnamefont{A.}~\bibnamefont{Magiera}},
  \bibinfo{author}{\bibfnamefont{A.}~\bibnamefont{Strza\l{}kowski}},
  \bibnamefont{and} \bibinfo{author}{\bibfnamefont{G.}~\bibnamefont{Willim}},
  \bibinfo{journal}{Nucl. Phys. A} \textbf{\bibinfo{volume}{368}},
  \bibinfo{pages}{173} (\bibinfo{year}{1981}).

\bibitem[{\citenamefont{Tricoire et~al.}(1982)\citenamefont{Tricoire, Gerschel,
  Perrin, Sergolle, Valentin, Bachelier, Doubre, and Gizon}}]{Tricoire82}
\bibinfo{author}{\bibfnamefont{H.}~\bibnamefont{Tricoire}},
  \bibinfo{author}{\bibfnamefont{C.}~\bibnamefont{Gerschel}},
  \bibinfo{author}{\bibfnamefont{N.}~\bibnamefont{Perrin}},
  \bibinfo{author}{\bibfnamefont{H.}~\bibnamefont{Sergolle}},
  \bibinfo{author}{\bibfnamefont{L.}~\bibnamefont{Valentin}},
  \bibinfo{author}{\bibfnamefont{D.}~\bibnamefont{Bachelier}},
  \bibinfo{author}{\bibfnamefont{H.}~\bibnamefont{Doubre}}, \bibnamefont{and}
  \bibinfo{author}{\bibfnamefont{J.}~\bibnamefont{Gizon}}, \bibinfo{journal}{Z.
  Phys. A} \textbf{\bibinfo{volume}{306}}, \bibinfo{pages}{127}
  (\bibinfo{year}{1982}).

\bibitem[{\citenamefont{Balster et~al.}(1987)\citenamefont{Balster, Crouzen,
  Goldhoorn, Siemssen, and Wilschut}}]{Balster87a}
\bibinfo{author}{\bibfnamefont{G.~J.} \bibnamefont{Balster}},
  \bibinfo{author}{\bibfnamefont{P.~C.~N.} \bibnamefont{Crouzen}},
  \bibinfo{author}{\bibfnamefont{P.~B.} \bibnamefont{Goldhoorn}},
  \bibinfo{author}{\bibfnamefont{R.~H.} \bibnamefont{Siemssen}},
  \bibnamefont{and} \bibinfo{author}{\bibfnamefont{H.~W.}
  \bibnamefont{Wilschut}}, \bibinfo{journal}{Nucl. Phys. A}
  \textbf{\bibinfo{volume}{468}}, \bibinfo{pages}{93} (\bibinfo{year}{1987}).

\bibitem[{\citenamefont{Parker et~al.}(1987)\citenamefont{Parker, Hogan, and
  Asher}}]{Parker87}
\bibinfo{author}{\bibfnamefont{D.~J.} \bibnamefont{Parker}},
  \bibinfo{author}{\bibfnamefont{J.~J.} \bibnamefont{Hogan}}, \bibnamefont{and}
  \bibinfo{author}{\bibfnamefont{J.}~\bibnamefont{Asher}},
  \bibinfo{journal}{Phys. Rev. C} \textbf{\bibinfo{volume}{35}},
  \bibinfo{pages}{161} (\bibinfo{year}{1987}).

\bibitem[{\citenamefont{{Zhang Li} et~al.}(1988)\citenamefont{{Zhang Li}, {Wang
  Da-yan}, {Jin Gen-ming}, {Zhang Bao-guo}, and {Wang Xi-ming}}}]{Li88}
\bibinfo{author}{\bibnamefont{{Zhang Li}}}, \bibinfo{author}{\bibnamefont{{Wang
  Da-yan}}}, \bibinfo{author}{\bibnamefont{{Jin Gen-ming}}},
  \bibinfo{author}{\bibnamefont{{Zhang Bao-guo}}}, \bibnamefont{and}
  \bibinfo{author}{\bibnamefont{{Wang Xi-ming}}}, \bibinfo{journal}{Phys. Rev.
  C} \textbf{\bibinfo{volume}{37}}, \bibinfo{pages}{669}
  (\bibinfo{year}{1988}).

\bibitem[{\citenamefont{Gavron}(1980)}]{Gavr80}
\bibinfo{author}{\bibfnamefont{A.}~\bibnamefont{Gavron}},
  \bibinfo{journal}{Phys. Rev. C} \textbf{\bibinfo{volume}{21}},
  \bibinfo{pages}{230} (\bibinfo{year}{1980}).

\bibitem[{\citenamefont{Santra et~al.}(2001)\citenamefont{Santra, Singh,
  Kailas, Chatterjee, Shrivastava, and Mahata}}]{Santra01}
\bibinfo{author}{\bibfnamefont{S.}~\bibnamefont{Santra}},
  \bibinfo{author}{\bibfnamefont{P.}~\bibnamefont{Singh}},
  \bibinfo{author}{\bibfnamefont{S.}~\bibnamefont{Kailas}},
  \bibinfo{author}{\bibfnamefont{A.}~\bibnamefont{Chatterjee}},
  \bibinfo{author}{\bibfnamefont{A.}~\bibnamefont{Shrivastava}},
  \bibnamefont{and} \bibinfo{author}{\bibfnamefont{K.}~\bibnamefont{Mahata}},
  \bibinfo{journal}{Phys. Rev. C} \textbf{\bibinfo{volume}{64}},
  \bibinfo{pages}{024602} (\bibinfo{year}{2001}).

\bibitem[{\citenamefont{Mukherjee et~al.}(2007)\citenamefont{Mukherjee, Hinde,
  Dasgupta, Hagino, Newton, and Butt}}]{Mukherjee07}
\bibinfo{author}{\bibfnamefont{A.}~\bibnamefont{Mukherjee}},
  \bibinfo{author}{\bibfnamefont{D.~J.} \bibnamefont{Hinde}},
  \bibinfo{author}{\bibfnamefont{M.}~\bibnamefont{Dasgupta}},
  \bibinfo{author}{\bibfnamefont{K.}~\bibnamefont{Hagino}},
  \bibinfo{author}{\bibfnamefont{J.~O.} \bibnamefont{Newton}},
  \bibnamefont{and} \bibinfo{author}{\bibfnamefont{R.~D.} \bibnamefont{Butt}},
  \bibinfo{journal}{Phys. Rev. C} \textbf{\bibinfo{volume}{75}},
  \bibinfo{pages}{044608} (\bibinfo{year}{2007}).

\bibitem[{\citenamefont{{F. A. Souza} et~al.}(2009)\citenamefont{{F. A. Souza},
  Beck, Carlin, Keeley, {R. Liguori Neto}, {M. M. de Moura}, Munhoz, {M. G. Del
  Santo}, {A. A. P. Suaide}, Szanto et~al.}}]{Souz09}
\bibinfo{author}{\bibnamefont{{F. A. Souza}}},
  \bibinfo{author}{\bibfnamefont{C.}~\bibnamefont{Beck}},
  \bibinfo{author}{\bibfnamefont{N.}~\bibnamefont{Carlin}},
  \bibinfo{author}{\bibfnamefont{N.}~\bibnamefont{Keeley}},
  \bibinfo{author}{\bibnamefont{{R. Liguori Neto}}},
  \bibinfo{author}{\bibnamefont{{M. M. de Moura}}},
  \bibinfo{author}{\bibfnamefont{M.~G.} \bibnamefont{Munhoz}},
  \bibinfo{author}{\bibnamefont{{M. G. Del Santo}}},
  \bibinfo{author}{\bibnamefont{{A. A. P. Suaide}}},
  \bibinfo{author}{\bibfnamefont{E.~M.} \bibnamefont{Szanto}},
  \bibnamefont{et~al.}, \bibinfo{journal}{Nucl. Phys. A}
  \textbf{\bibinfo{volume}{821}}, \bibinfo{pages}{36} (\bibinfo{year}{2009}).

\bibitem[{\citenamefont{Shrivastava et~al.}(2006)\citenamefont{Shrivastava,
  Navin, Keeley, Mahata, Ramachandran, Nanal, Parkar, Chatterjee, and
  Kailas}}]{Shri06}
\bibinfo{author}{\bibfnamefont{A.}~\bibnamefont{Shrivastava}},
  \bibinfo{author}{\bibfnamefont{A.}~\bibnamefont{Navin}},
  \bibinfo{author}{\bibfnamefont{N.}~\bibnamefont{Keeley}},
  \bibinfo{author}{\bibfnamefont{K.}~\bibnamefont{Mahata}},
  \bibinfo{author}{\bibfnamefont{K.}~\bibnamefont{Ramachandran}},
  \bibinfo{author}{\bibfnamefont{V.}~\bibnamefont{Nanal}},
  \bibinfo{author}{\bibfnamefont{V.~V.} \bibnamefont{Parkar}},
  \bibinfo{author}{\bibfnamefont{A.}~\bibnamefont{Chatterjee}},
  \bibnamefont{and} \bibinfo{author}{\bibfnamefont{S.}~\bibnamefont{Kailas}},
  \bibinfo{journal}{Phys. Lett. B} \textbf{\bibinfo{volume}{633}},
  \bibinfo{pages}{463} (\bibinfo{year}{2006}).

\bibitem[{\citenamefont{Chattopadhyay et~al.}(2016)\citenamefont{Chattopadhyay,
  Santra, Pal, Kundu, Ramachandran, Tripathi, Sarkar, Sodaye, Nayak, Saxena
  et~al.}}]{Chatto16}
\bibinfo{author}{\bibfnamefont{D.}~\bibnamefont{Chattopadhyay}},
  \bibinfo{author}{\bibfnamefont{S.}~\bibnamefont{Santra}},
  \bibinfo{author}{\bibfnamefont{A.}~\bibnamefont{Pal}},
  \bibinfo{author}{\bibfnamefont{A.}~\bibnamefont{Kundu}},
  \bibinfo{author}{\bibfnamefont{K.}~\bibnamefont{Ramachandran}},
  \bibinfo{author}{\bibfnamefont{R.}~\bibnamefont{Tripathi}},
  \bibinfo{author}{\bibfnamefont{D.}~\bibnamefont{Sarkar}},
  \bibinfo{author}{\bibfnamefont{S.}~\bibnamefont{Sodaye}},
  \bibinfo{author}{\bibfnamefont{B.~K.} \bibnamefont{Nayak}},
  \bibinfo{author}{\bibfnamefont{A.}~\bibnamefont{Saxena}},
  \bibnamefont{et~al.}, \bibinfo{journal}{Phys. Rev. C}
  \textbf{\bibinfo{volume}{94}}, \bibinfo{pages}{061602(R)}
  (\bibinfo{year}{2016}).

\bibitem[{\citenamefont{Pradhan et~al.}(2013)\citenamefont{Pradhan, Mukherjee,
  {Subinit Roy}, Basu, Goswami, Kshetri, Palit, Parkar, Ray, {M. Saha Sarkar}
  et~al.}}]{Prad13}
\bibinfo{author}{\bibfnamefont{M.~K.} \bibnamefont{Pradhan}},
  \bibinfo{author}{\bibfnamefont{A.}~\bibnamefont{Mukherjee}},
  \bibinfo{author}{\bibnamefont{{Subinit Roy}}},
  \bibinfo{author}{\bibfnamefont{P.}~\bibnamefont{Basu}},
  \bibinfo{author}{\bibfnamefont{A.}~\bibnamefont{Goswami}},
  \bibinfo{author}{\bibfnamefont{R.}~\bibnamefont{Kshetri}},
  \bibinfo{author}{\bibfnamefont{R.}~\bibnamefont{Palit}},
  \bibinfo{author}{\bibfnamefont{V.~V.} \bibnamefont{Parkar}},
  \bibinfo{author}{\bibfnamefont{M.}~\bibnamefont{Ray}},
  \bibinfo{author}{\bibnamefont{{M. Saha Sarkar}}}, \bibnamefont{et~al.},
  \bibinfo{journal}{Phys. Rev. C} \textbf{\bibinfo{volume}{88}},
  \bibinfo{pages}{064603} (\bibinfo{year}{2013}).

\bibitem[{\citenamefont{Ost et~al.}(1972)\citenamefont{Ost, Speth, Pfeiffer,
  and Bethge}}]{Ost72}
\bibinfo{author}{\bibfnamefont{R.}~\bibnamefont{Ost}},
  \bibinfo{author}{\bibfnamefont{E.}~\bibnamefont{Speth}},
  \bibinfo{author}{\bibfnamefont{K.~O.} \bibnamefont{Pfeiffer}},
  \bibnamefont{and} \bibinfo{author}{\bibfnamefont{K.}~\bibnamefont{Bethge}},
  \bibinfo{journal}{Phys. Rev. C} \textbf{\bibinfo{volume}{5}},
  \bibinfo{pages}{1835} (\bibinfo{year}{1972}).

\bibitem[{\citenamefont{Signorini et~al.}(2001)\citenamefont{Signorini,
  Mazzocco, Prete, Soramel, Stroe, Andrighetto, Thompson, Vitturi, Brondi,
  Cinausero et~al.}}]{Sign01}
\bibinfo{author}{\bibfnamefont{C.}~\bibnamefont{Signorini}},
  \bibinfo{author}{\bibfnamefont{M.}~\bibnamefont{Mazzocco}},
  \bibinfo{author}{\bibfnamefont{G.~F.} \bibnamefont{Prete}},
  \bibinfo{author}{\bibfnamefont{F.}~\bibnamefont{Soramel}},
  \bibinfo{author}{\bibfnamefont{L.}~\bibnamefont{Stroe}},
  \bibinfo{author}{\bibfnamefont{A.}~\bibnamefont{Andrighetto}},
  \bibinfo{author}{\bibfnamefont{I.~J.} \bibnamefont{Thompson}},
  \bibinfo{author}{\bibfnamefont{A.}~\bibnamefont{Vitturi}},
  \bibinfo{author}{\bibfnamefont{A.}~\bibnamefont{Brondi}},
  \bibinfo{author}{\bibfnamefont{M.}~\bibnamefont{Cinausero}},
  \bibnamefont{et~al.}, \bibinfo{journal}{Eur. Phys. J. A}
  \textbf{\bibinfo{volume}{10}}, \bibinfo{pages}{249} (\bibinfo{year}{2001}).

\bibitem[{\citenamefont{Pakou et~al.}(2005)\citenamefont{Pakou, Nicolis, Rusek,
  Alamanos, Doukelis, Gillibert, Kalyva, Kokkoris, Lagoyannis, Musumarra
  et~al.}}]{Pako05}
\bibinfo{author}{\bibfnamefont{A.}~\bibnamefont{Pakou}},
  \bibinfo{author}{\bibfnamefont{N.~G.} \bibnamefont{Nicolis}},
  \bibinfo{author}{\bibfnamefont{K.}~\bibnamefont{Rusek}},
  \bibinfo{author}{\bibfnamefont{N.}~\bibnamefont{Alamanos}},
  \bibinfo{author}{\bibfnamefont{G.}~\bibnamefont{Doukelis}},
  \bibinfo{author}{\bibfnamefont{A.}~\bibnamefont{Gillibert}},
  \bibinfo{author}{\bibfnamefont{G.}~\bibnamefont{Kalyva}},
  \bibinfo{author}{\bibfnamefont{M.}~\bibnamefont{Kokkoris}},
  \bibinfo{author}{\bibfnamefont{A.}~\bibnamefont{Lagoyannis}},
  \bibinfo{author}{\bibfnamefont{A.}~\bibnamefont{Musumarra}},
  \bibnamefont{et~al.}, \bibinfo{journal}{Phys. Rev. C}
  \textbf{\bibinfo{volume}{71}}, \bibinfo{pages}{064602}
  (\bibinfo{year}{2005}).

\bibitem[{\citenamefont{Souza et~al.}(2010)\citenamefont{Souza, Carlin, Beck,
  Keeley, Diaz-Torres, {R. Liguori Neto}, Siqueira-Mello, {M. M. de Moura},
  Munhoz, Oliveira et~al.}}]{Souza2010}
\bibinfo{author}{\bibfnamefont{F.~A.} \bibnamefont{Souza}},
  \bibinfo{author}{\bibfnamefont{N.}~\bibnamefont{Carlin}},
  \bibinfo{author}{\bibfnamefont{C.}~\bibnamefont{Beck}},
  \bibinfo{author}{\bibfnamefont{N.}~\bibnamefont{Keeley}},
  \bibinfo{author}{\bibfnamefont{A.}~\bibnamefont{Diaz-Torres}},
  \bibinfo{author}{\bibnamefont{{R. Liguori Neto}}},
  \bibinfo{author}{\bibfnamefont{C.}~\bibnamefont{Siqueira-Mello}},
  \bibinfo{author}{\bibnamefont{{M. M. de Moura}}},
  \bibinfo{author}{\bibfnamefont{M.~G.} \bibnamefont{Munhoz}},
  \bibinfo{author}{\bibfnamefont{R.~A.~N.} \bibnamefont{Oliveira}},
  \bibnamefont{et~al.}, \bibinfo{journal}{Eur. Phys. J. A}
  \textbf{\bibinfo{volume}{44}}, \bibinfo{pages}{181} (\bibinfo{year}{2010}).

\bibitem[{\citenamefont{Pradhan et~al.}(2011)\citenamefont{Pradhan, Mukherjee,
  Basu, Goswami, Kshetri, {Subinit Roy}, {P. Roy Chowdhury}, {M. Saha Sarkar},
  Palit, Parkar et~al.}}]{Pradhan11}
\bibinfo{author}{\bibfnamefont{M.~K.} \bibnamefont{Pradhan}},
  \bibinfo{author}{\bibfnamefont{A.}~\bibnamefont{Mukherjee}},
  \bibinfo{author}{\bibfnamefont{P.}~\bibnamefont{Basu}},
  \bibinfo{author}{\bibfnamefont{A.}~\bibnamefont{Goswami}},
  \bibinfo{author}{\bibfnamefont{R.}~\bibnamefont{Kshetri}},
  \bibinfo{author}{\bibnamefont{{Subinit Roy}}},
  \bibinfo{author}{\bibnamefont{{P. Roy Chowdhury}}},
  \bibinfo{author}{\bibnamefont{{M. Saha Sarkar}}},
  \bibinfo{author}{\bibfnamefont{R.}~\bibnamefont{Palit}},
  \bibinfo{author}{\bibfnamefont{V.~V.} \bibnamefont{Parkar}},
  \bibnamefont{et~al.}, \bibinfo{journal}{Phys. Rev. C}
  \textbf{\bibinfo{volume}{83}}, \bibinfo{pages}{064606}
  (\bibinfo{year}{2011}).

\bibitem[{\citenamefont{Pandit et~al.}(2016)\citenamefont{Pandit, Shrivastava,
  Mahata, Keeley, Parkar, Rout, Ramachandran, Martel, Palshetkar, Kumar
  et~al.}}]{Pand16}
\bibinfo{author}{\bibfnamefont{S.~K.} \bibnamefont{Pandit}},
  \bibinfo{author}{\bibfnamefont{A.}~\bibnamefont{Shrivastava}},
  \bibinfo{author}{\bibfnamefont{K.}~\bibnamefont{Mahata}},
  \bibinfo{author}{\bibfnamefont{N.}~\bibnamefont{Keeley}},
  \bibinfo{author}{\bibfnamefont{V.~V.} \bibnamefont{Parkar}},
  \bibinfo{author}{\bibfnamefont{P.~C.} \bibnamefont{Rout}},
  \bibinfo{author}{\bibfnamefont{K.}~\bibnamefont{Ramachandran}},
  \bibinfo{author}{\bibfnamefont{I.}~\bibnamefont{Martel}},
  \bibinfo{author}{\bibfnamefont{C.~S.} \bibnamefont{Palshetkar}},
  \bibinfo{author}{\bibfnamefont{A.}~\bibnamefont{Kumar}},
  \bibnamefont{et~al.}, \bibinfo{journal}{Phys. Rev. C}
  \textbf{\bibinfo{volume}{93}}, \bibinfo{pages}{061602(R)}
  (\bibinfo{year}{2016}).

\bibitem[{\citenamefont{Chattopadhyay
  et~al.}(2018{\natexlab{a}})\citenamefont{Chattopadhyay, Santra, Pal, Kundu,
  Ramachandran, Tripathi, Roy, Nag, Sawant, Nayak et~al.}}]{Chattopadhyay2018}
\bibinfo{author}{\bibfnamefont{D.}~\bibnamefont{Chattopadhyay}},
  \bibinfo{author}{\bibfnamefont{S.}~\bibnamefont{Santra}},
  \bibinfo{author}{\bibfnamefont{A.}~\bibnamefont{Pal}},
  \bibinfo{author}{\bibfnamefont{A.}~\bibnamefont{Kundu}},
  \bibinfo{author}{\bibfnamefont{K.}~\bibnamefont{Ramachandran}},
  \bibinfo{author}{\bibfnamefont{R.}~\bibnamefont{Tripathi}},
  \bibinfo{author}{\bibfnamefont{B.~J.} \bibnamefont{Roy}},
  \bibinfo{author}{\bibfnamefont{T.~N.} \bibnamefont{Nag}},
  \bibinfo{author}{\bibfnamefont{Y.}~\bibnamefont{Sawant}},
  \bibinfo{author}{\bibfnamefont{B.~K.} \bibnamefont{Nayak}},
  \bibnamefont{et~al.}, \bibinfo{journal}{Phys. Rev. C}
  \textbf{\bibinfo{volume}{97}}, \bibinfo{pages}{051601(R)}
  (\bibinfo{year}{2018}{\natexlab{a}}).

\bibitem[{\citenamefont{Chattopadhyay
  et~al.}(2018{\natexlab{b}})\citenamefont{Chattopadhyay, Santra, Pal, Kundu,
  Ramachandran, Tripathi, Roy, Sawant, Nayak, Saxena et~al.}}]{Chatto18b}
\bibinfo{author}{\bibfnamefont{D.}~\bibnamefont{Chattopadhyay}},
  \bibinfo{author}{\bibfnamefont{S.}~\bibnamefont{Santra}},
  \bibinfo{author}{\bibfnamefont{A.}~\bibnamefont{Pal}},
  \bibinfo{author}{\bibfnamefont{A.}~\bibnamefont{Kundu}},
  \bibinfo{author}{\bibfnamefont{K.}~\bibnamefont{Ramachandran}},
  \bibinfo{author}{\bibfnamefont{R.}~\bibnamefont{Tripathi}},
  \bibinfo{author}{\bibfnamefont{B.~J.} \bibnamefont{Roy}},
  \bibinfo{author}{\bibfnamefont{Y.}~\bibnamefont{Sawant}},
  \bibinfo{author}{\bibfnamefont{B.~K.} \bibnamefont{Nayak}},
  \bibinfo{author}{\bibfnamefont{A.}~\bibnamefont{Saxena}},
  \bibnamefont{et~al.}, \bibinfo{journal}{Phys. Rev. C}
  \textbf{\bibinfo{volume}{98}}, \bibinfo{pages}{014609}
  (\bibinfo{year}{2018}{\natexlab{b}}).

\bibitem[{\citenamefont{Chamon et~al.}(2002)\citenamefont{Chamon, Carlson,
  Gasques, Pereira, De~Conti, Alvarez, Hussein, C\^andido~Ribeiro, Rossi, and
  Silva}}]{Chamon02}
\bibinfo{author}{\bibfnamefont{L.~C.} \bibnamefont{Chamon}},
  \bibinfo{author}{\bibfnamefont{B.~V.} \bibnamefont{Carlson}},
  \bibinfo{author}{\bibfnamefont{L.~R.} \bibnamefont{Gasques}},
  \bibinfo{author}{\bibfnamefont{D.}~\bibnamefont{Pereira}},
  \bibinfo{author}{\bibfnamefont{C.}~\bibnamefont{De~Conti}},
  \bibinfo{author}{\bibfnamefont{M.~A.~G.} \bibnamefont{Alvarez}},
  \bibinfo{author}{\bibfnamefont{M.~S.} \bibnamefont{Hussein}},
  \bibinfo{author}{\bibfnamefont{M.~A.} \bibnamefont{C\^andido~Ribeiro}},
  \bibinfo{author}{\bibfnamefont{E.~S.} \bibnamefont{Rossi}}, \bibnamefont{and}
  \bibinfo{author}{\bibfnamefont{C.~P.} \bibnamefont{Silva}},
  \bibinfo{journal}{Phys. Rev. C} \textbf{\bibinfo{volume}{66}},
  \bibinfo{pages}{014610} (\bibinfo{year}{2002}).

\bibitem[{\citenamefont{Palshetkar
  et~al.}(2014{\natexlab{a}})\citenamefont{Palshetkar, {Shital Thakur}, Nanal,
  Shrivastava, Dokania, Singh, Parkar, Rout, Palit, Pillay et~al.}}]{Pals14}
\bibinfo{author}{\bibfnamefont{C.~S.} \bibnamefont{Palshetkar}},
  \bibinfo{author}{\bibnamefont{{Shital Thakur}}},
  \bibinfo{author}{\bibfnamefont{V.}~\bibnamefont{Nanal}},
  \bibinfo{author}{\bibfnamefont{A.}~\bibnamefont{Shrivastava}},
  \bibinfo{author}{\bibfnamefont{N.}~\bibnamefont{Dokania}},
  \bibinfo{author}{\bibfnamefont{V.}~\bibnamefont{Singh}},
  \bibinfo{author}{\bibfnamefont{V.~V.} \bibnamefont{Parkar}},
  \bibinfo{author}{\bibfnamefont{P.~C.} \bibnamefont{Rout}},
  \bibinfo{author}{\bibfnamefont{R.}~\bibnamefont{Palit}},
  \bibinfo{author}{\bibfnamefont{R.~G.} \bibnamefont{Pillay}},
  \bibnamefont{et~al.}, \bibinfo{journal}{Phys. Rev. C}
  \textbf{\bibinfo{volume}{89}}, \bibinfo{pages}{024607}
  (\bibinfo{year}{2014}{\natexlab{a}}).

\bibitem[{\citenamefont{Shrivastava et~al.}(2009)\citenamefont{Shrivastava,
  Navin, Lemasson, Ramachandran, Nanal, Rejmund, Hagino, Ichikawa,
  Bhattacharyya, Chatterjee et~al.}}]{Shrivastava09}
\bibinfo{author}{\bibfnamefont{A.}~\bibnamefont{Shrivastava}},
  \bibinfo{author}{\bibfnamefont{A.}~\bibnamefont{Navin}},
  \bibinfo{author}{\bibfnamefont{A.}~\bibnamefont{Lemasson}},
  \bibinfo{author}{\bibfnamefont{K.}~\bibnamefont{Ramachandran}},
  \bibinfo{author}{\bibfnamefont{V.}~\bibnamefont{Nanal}},
  \bibinfo{author}{\bibfnamefont{M.}~\bibnamefont{Rejmund}},
  \bibinfo{author}{\bibfnamefont{K.}~\bibnamefont{Hagino}},
  \bibinfo{author}{\bibfnamefont{T.}~\bibnamefont{Ichikawa}},
  \bibinfo{author}{\bibfnamefont{S.}~\bibnamefont{Bhattacharyya}},
  \bibinfo{author}{\bibfnamefont{A.}~\bibnamefont{Chatterjee}},
  \bibnamefont{et~al.}, \bibinfo{journal}{Phys. Rev. Lett.}
  \textbf{\bibinfo{volume}{103}}, \bibinfo{pages}{232702}
  (\bibinfo{year}{2009}).

\bibitem[{\citenamefont{Broda et~al.}(1975)\citenamefont{Broda, Ishihara,
  Herskind, Oeschler, Ogaza, and Ryde}}]{Brod75}
\bibinfo{author}{\bibfnamefont{R.}~\bibnamefont{Broda}},
  \bibinfo{author}{\bibfnamefont{M.}~\bibnamefont{Ishihara}},
  \bibinfo{author}{\bibfnamefont{B.}~\bibnamefont{Herskind}},
  \bibinfo{author}{\bibfnamefont{H.}~\bibnamefont{Oeschler}},
  \bibinfo{author}{\bibfnamefont{S.}~\bibnamefont{Ogaza}}, \bibnamefont{and}
  \bibinfo{author}{\bibfnamefont{H.}~\bibnamefont{Ryde}},
  \bibinfo{journal}{Nucl. Phys. A} \textbf{\bibinfo{volume}{248}},
  \bibinfo{pages}{356} (\bibinfo{year}{1975}).

\bibitem[{\citenamefont{Shrivastava et~al.}(2013)\citenamefont{Shrivastava,
  Navin, Diaz-Torres, Nanal, Ramachandran, Rejmund, Bhattacharyya, Chatterjee,
  Kailas, Lemasson et~al.}}]{Ara13}
\bibinfo{author}{\bibfnamefont{A.}~\bibnamefont{Shrivastava}},
  \bibinfo{author}{\bibfnamefont{A.}~\bibnamefont{Navin}},
  \bibinfo{author}{\bibfnamefont{A.}~\bibnamefont{Diaz-Torres}},
  \bibinfo{author}{\bibfnamefont{V.}~\bibnamefont{Nanal}},
  \bibinfo{author}{\bibfnamefont{K.}~\bibnamefont{Ramachandran}},
  \bibinfo{author}{\bibfnamefont{M.}~\bibnamefont{Rejmund}},
  \bibinfo{author}{\bibfnamefont{S.}~\bibnamefont{Bhattacharyya}},
  \bibinfo{author}{\bibfnamefont{A.}~\bibnamefont{Chatterjee}},
  \bibinfo{author}{\bibfnamefont{S.}~\bibnamefont{Kailas}},
  \bibinfo{author}{\bibfnamefont{A.}~\bibnamefont{Lemasson}},
  \bibnamefont{et~al.}, \bibinfo{journal}{Phys. Lett. B}
  \textbf{\bibinfo{volume}{718}}, \bibinfo{pages}{931} (\bibinfo{year}{2013}).

\bibitem[{\citenamefont{Dasgupta et~al.}(2004)\citenamefont{Dasgupta, Gomes,
  Hinde, Moraes, Anjos, Berriman, Butt, Carlin, Lubian, Morton
  et~al.}}]{Dasgupta04}
\bibinfo{author}{\bibfnamefont{M.}~\bibnamefont{Dasgupta}},
  \bibinfo{author}{\bibfnamefont{P.~R.~S.} \bibnamefont{Gomes}},
  \bibinfo{author}{\bibfnamefont{D.~J.} \bibnamefont{Hinde}},
  \bibinfo{author}{\bibfnamefont{S.~B.} \bibnamefont{Moraes}},
  \bibinfo{author}{\bibfnamefont{R.~M.} \bibnamefont{Anjos}},
  \bibinfo{author}{\bibfnamefont{A.~C.} \bibnamefont{Berriman}},
  \bibinfo{author}{\bibfnamefont{R.~D.} \bibnamefont{Butt}},
  \bibinfo{author}{\bibfnamefont{N.}~\bibnamefont{Carlin}},
  \bibinfo{author}{\bibfnamefont{J.}~\bibnamefont{Lubian}},
  \bibinfo{author}{\bibfnamefont{C.~R.} \bibnamefont{Morton}},
  \bibnamefont{et~al.}, \bibinfo{journal}{Phys. Rev. C}
  \textbf{\bibinfo{volume}{70}}, \bibinfo{pages}{024606}
  (\bibinfo{year}{2004}).

\bibitem[{\citenamefont{Pakou et~al.}(2006)\citenamefont{Pakou, Alamanos,
  Clarke, Davis, Doukelis, Kalyva, Kokkoris, Lagoyannis, Mertzimekis, Musumarra
  et~al.}}]{Pako06}
\bibinfo{author}{\bibfnamefont{A.}~\bibnamefont{Pakou}},
  \bibinfo{author}{\bibfnamefont{N.}~\bibnamefont{Alamanos}},
  \bibinfo{author}{\bibfnamefont{N.~M.} \bibnamefont{Clarke}},
  \bibinfo{author}{\bibfnamefont{N.~J.} \bibnamefont{Davis}},
  \bibinfo{author}{\bibfnamefont{G.}~\bibnamefont{Doukelis}},
  \bibinfo{author}{\bibfnamefont{G.}~\bibnamefont{Kalyva}},
  \bibinfo{author}{\bibfnamefont{M.}~\bibnamefont{Kokkoris}},
  \bibinfo{author}{\bibfnamefont{A.}~\bibnamefont{Lagoyannis}},
  \bibinfo{author}{\bibfnamefont{T.~J.} \bibnamefont{Mertzimekis}},
  \bibinfo{author}{\bibfnamefont{A.}~\bibnamefont{Musumarra}},
  \bibnamefont{et~al.}, \bibinfo{journal}{Phys. Lett. B}
  \textbf{\bibinfo{volume}{633}}, \bibinfo{pages}{691} (\bibinfo{year}{2006}).

\bibitem[{\citenamefont{Signorini et~al.}(2003)\citenamefont{Signorini,
  Edifizi, Mazzocco, Lunardon, Fabris, Vitturi, Scopel, Soramel, Stroe, Prete
  et~al.}}]{Sign03}
\bibinfo{author}{\bibfnamefont{C.}~\bibnamefont{Signorini}},
  \bibinfo{author}{\bibfnamefont{A.}~\bibnamefont{Edifizi}},
  \bibinfo{author}{\bibfnamefont{M.}~\bibnamefont{Mazzocco}},
  \bibinfo{author}{\bibfnamefont{M.}~\bibnamefont{Lunardon}},
  \bibinfo{author}{\bibfnamefont{D.}~\bibnamefont{Fabris}},
  \bibinfo{author}{\bibfnamefont{A.}~\bibnamefont{Vitturi}},
  \bibinfo{author}{\bibfnamefont{P.}~\bibnamefont{Scopel}},
  \bibinfo{author}{\bibfnamefont{F.}~\bibnamefont{Soramel}},
  \bibinfo{author}{\bibfnamefont{L.}~\bibnamefont{Stroe}},
  \bibinfo{author}{\bibfnamefont{G.}~\bibnamefont{Prete}},
  \bibnamefont{et~al.}, \bibinfo{journal}{Phys. Rev. C}
  \textbf{\bibinfo{volume}{67}}, \bibinfo{pages}{044607}
  (\bibinfo{year}{2003}).

\bibitem[{\citenamefont{Santra et~al.}(2009)\citenamefont{Santra, Parkar,
  Ramachandran, Pal, Shrivastava, Roy, Nayak, Chatterjee, Choudhury, and
  Kailas}}]{Sant09}
\bibinfo{author}{\bibfnamefont{S.}~\bibnamefont{Santra}},
  \bibinfo{author}{\bibfnamefont{V.~V.} \bibnamefont{Parkar}},
  \bibinfo{author}{\bibfnamefont{K.}~\bibnamefont{Ramachandran}},
  \bibinfo{author}{\bibfnamefont{U.~K.} \bibnamefont{Pal}},
  \bibinfo{author}{\bibfnamefont{A.}~\bibnamefont{Shrivastava}},
  \bibinfo{author}{\bibfnamefont{B.~J.} \bibnamefont{Roy}},
  \bibinfo{author}{\bibfnamefont{B.~K.} \bibnamefont{Nayak}},
  \bibinfo{author}{\bibfnamefont{A.}~\bibnamefont{Chatterjee}},
  \bibinfo{author}{\bibfnamefont{R.~K.} \bibnamefont{Choudhury}},
  \bibnamefont{and} \bibinfo{author}{\bibfnamefont{S.}~\bibnamefont{Kailas}},
  \bibinfo{journal}{Phys. Lett. B} \textbf{\bibinfo{volume}{677}},
  \bibinfo{pages}{139} (\bibinfo{year}{2009}).

\bibitem[{\citenamefont{Signorini et~al.}(2004)\citenamefont{Signorini,
  Glodariu, Liu, Mazzocco, Ruan, and Soramel}}]{Sig04}
\bibinfo{author}{\bibfnamefont{C.}~\bibnamefont{Signorini}},
  \bibinfo{author}{\bibfnamefont{T.}~\bibnamefont{Glodariu}},
  \bibinfo{author}{\bibfnamefont{Z.~H.} \bibnamefont{Liu}},
  \bibinfo{author}{\bibfnamefont{M.}~\bibnamefont{Mazzocco}},
  \bibinfo{author}{\bibfnamefont{M.}~\bibnamefont{Ruan}}, \bibnamefont{and}
  \bibinfo{author}{\bibfnamefont{F.}~\bibnamefont{Soramel}},
  \bibinfo{journal}{Progress of Theoretical Physics Supplement}
  \textbf{\bibinfo{volume}{154}}, \bibinfo{pages}{272} (\bibinfo{year}{2004}).

\bibitem[{\citenamefont{Woolliscroft et~al.}(2003)\citenamefont{Woolliscroft,
  Clarke, Fulton, Cowin, Dasgupta, Hinde, Morton, and Berriman}}]{Wooll03}
\bibinfo{author}{\bibfnamefont{R.~J.} \bibnamefont{Woolliscroft}},
  \bibinfo{author}{\bibfnamefont{N.~M.} \bibnamefont{Clarke}},
  \bibinfo{author}{\bibfnamefont{B.~R.} \bibnamefont{Fulton}},
  \bibinfo{author}{\bibfnamefont{R.~L.} \bibnamefont{Cowin}},
  \bibinfo{author}{\bibfnamefont{M.}~\bibnamefont{Dasgupta}},
  \bibinfo{author}{\bibfnamefont{D.~J.} \bibnamefont{Hinde}},
  \bibinfo{author}{\bibfnamefont{C.~R.} \bibnamefont{Morton}},
  \bibnamefont{and} \bibinfo{author}{\bibfnamefont{A.~C.}
  \bibnamefont{Berriman}}, \bibinfo{journal}{Phys. Rev. C}
  \textbf{\bibinfo{volume}{68}}, \bibinfo{pages}{014611}
  (\bibinfo{year}{2003}).

\bibitem[{\citenamefont{Palshetkar
  et~al.}(2014{\natexlab{b}})\citenamefont{Palshetkar, Santra, Shrivastava,
  Chatterjee, Pandit, Ramachandran, Parkar, Nanal, Jha, Roy et~al.}}]{Pals14b}
\bibinfo{author}{\bibfnamefont{C.~S.} \bibnamefont{Palshetkar}},
  \bibinfo{author}{\bibfnamefont{S.}~\bibnamefont{Santra}},
  \bibinfo{author}{\bibfnamefont{A.}~\bibnamefont{Shrivastava}},
  \bibinfo{author}{\bibfnamefont{A.}~\bibnamefont{Chatterjee}},
  \bibinfo{author}{\bibfnamefont{S.~K.} \bibnamefont{Pandit}},
  \bibinfo{author}{\bibfnamefont{K.}~\bibnamefont{Ramachandran}},
  \bibinfo{author}{\bibfnamefont{V.~V.} \bibnamefont{Parkar}},
  \bibinfo{author}{\bibfnamefont{V.}~\bibnamefont{Nanal}},
  \bibinfo{author}{\bibfnamefont{V.}~\bibnamefont{Jha}},
  \bibinfo{author}{\bibfnamefont{B.~J.} \bibnamefont{Roy}},
  \bibnamefont{et~al.}, \bibinfo{journal}{Phys. Rev. C}
  \textbf{\bibinfo{volume}{89}}, \bibinfo{pages}{064610}
  (\bibinfo{year}{2014}{\natexlab{b}}).

\bibitem[{\citenamefont{Yu et~al.}(2010)\citenamefont{Yu, Zhang, Jia, Zhang,
  Ruan, Yang, Wu, Xu, and Bai}}]{Yu10}
\bibinfo{author}{\bibfnamefont{N.}~\bibnamefont{Yu}},
  \bibinfo{author}{\bibfnamefont{H.~Q.} \bibnamefont{Zhang}},
  \bibinfo{author}{\bibfnamefont{H.~M.} \bibnamefont{Jia}},
  \bibinfo{author}{\bibfnamefont{S.~T.} \bibnamefont{Zhang}},
  \bibinfo{author}{\bibfnamefont{M.}~\bibnamefont{Ruan}},
  \bibinfo{author}{\bibfnamefont{F.}~\bibnamefont{Yang}},
  \bibinfo{author}{\bibfnamefont{Z.~D.} \bibnamefont{Wu}},
  \bibinfo{author}{\bibfnamefont{X.~X.} \bibnamefont{Xu}}, \bibnamefont{and}
  \bibinfo{author}{\bibfnamefont{C.~L.} \bibnamefont{Bai}},
  \bibinfo{journal}{J. Phys. G : Nucl. Part. Phys.}
  \textbf{\bibinfo{volume}{37}}, \bibinfo{pages}{075108}
  (\bibinfo{year}{2010}).

\bibitem[{\citenamefont{Fang et~al.}(2015)\citenamefont{Fang, Gomes, Lubian,
  Liu, Zhou, {D. R. Mendes Junior}, Zhang, Zhang, Li, Wang et~al.}}]{Fang15}
\bibinfo{author}{\bibfnamefont{Y.~D.} \bibnamefont{Fang}},
  \bibinfo{author}{\bibfnamefont{P.~R.~S.} \bibnamefont{Gomes}},
  \bibinfo{author}{\bibfnamefont{J.}~\bibnamefont{Lubian}},
  \bibinfo{author}{\bibfnamefont{M.~L.} \bibnamefont{Liu}},
  \bibinfo{author}{\bibfnamefont{X.~H.} \bibnamefont{Zhou}},
  \bibinfo{author}{\bibnamefont{{D. R. Mendes Junior}}},
  \bibinfo{author}{\bibfnamefont{N.~T.} \bibnamefont{Zhang}},
  \bibinfo{author}{\bibfnamefont{Y.~H.} \bibnamefont{Zhang}},
  \bibinfo{author}{\bibfnamefont{G.~S.} \bibnamefont{Li}},
  \bibinfo{author}{\bibfnamefont{J.~G.} \bibnamefont{Wang}},
  \bibnamefont{et~al.}, \bibinfo{journal}{Phys. Rev. C}
  \textbf{\bibinfo{volume}{91}}, \bibinfo{pages}{014608}
  (\bibinfo{year}{2015}).

\bibitem[{\citenamefont{Jha et~al.}(2014)\citenamefont{Jha, Parkar, and
  Kailas}}]{Jha14}
\bibinfo{author}{\bibfnamefont{V.}~\bibnamefont{Jha}},
  \bibinfo{author}{\bibfnamefont{V.~V.} \bibnamefont{Parkar}},
  \bibnamefont{and} \bibinfo{author}{\bibfnamefont{S.}~\bibnamefont{Kailas}},
  \bibinfo{journal}{Phys. Rev. C} \textbf{\bibinfo{volume}{89}},
  \bibinfo{pages}{034605} (\bibinfo{year}{2014}).

\bibitem[{\citenamefont{Di~Pietro et~al.}(2004)\citenamefont{Di~Pietro,
  Figuera, Amorini, Angulo, Cardella, Cherubini, Davinson, Leanza, Lu, Mahmud
  et~al.}}]{Piet04}
\bibinfo{author}{\bibfnamefont{A.}~\bibnamefont{Di~Pietro}},
  \bibinfo{author}{\bibfnamefont{P.}~\bibnamefont{Figuera}},
  \bibinfo{author}{\bibfnamefont{F.}~\bibnamefont{Amorini}},
  \bibinfo{author}{\bibfnamefont{C.}~\bibnamefont{Angulo}},
  \bibinfo{author}{\bibfnamefont{G.}~\bibnamefont{Cardella}},
  \bibinfo{author}{\bibfnamefont{S.}~\bibnamefont{Cherubini}},
  \bibinfo{author}{\bibfnamefont{T.}~\bibnamefont{Davinson}},
  \bibinfo{author}{\bibfnamefont{D.}~\bibnamefont{Leanza}},
  \bibinfo{author}{\bibfnamefont{J.}~\bibnamefont{Lu}},
  \bibinfo{author}{\bibfnamefont{H.}~\bibnamefont{Mahmud}},
  \bibnamefont{et~al.}, \bibinfo{journal}{Phys. Rev. C}
  \textbf{\bibinfo{volume}{69}}, \bibinfo{pages}{044613}
  (\bibinfo{year}{2004}).

\bibitem[{\citenamefont{Scuderi et~al.}(2011)\citenamefont{Scuderi, Di~Pietro,
  Figuera, Fisichella, Amorini, Angulo, Cardella, Casarejos, Lattuada, Milin
  et~al.}}]{Scu11}
\bibinfo{author}{\bibfnamefont{V.}~\bibnamefont{Scuderi}},
  \bibinfo{author}{\bibfnamefont{A.}~\bibnamefont{Di~Pietro}},
  \bibinfo{author}{\bibfnamefont{P.}~\bibnamefont{Figuera}},
  \bibinfo{author}{\bibfnamefont{M.}~\bibnamefont{Fisichella}},
  \bibinfo{author}{\bibfnamefont{F.}~\bibnamefont{Amorini}},
  \bibinfo{author}{\bibfnamefont{C.}~\bibnamefont{Angulo}},
  \bibinfo{author}{\bibfnamefont{G.}~\bibnamefont{Cardella}},
  \bibinfo{author}{\bibfnamefont{E.}~\bibnamefont{Casarejos}},
  \bibinfo{author}{\bibfnamefont{M.}~\bibnamefont{Lattuada}},
  \bibinfo{author}{\bibfnamefont{M.}~\bibnamefont{Milin}},
  \bibnamefont{et~al.}, \bibinfo{journal}{Phys. Rev. C}
  \textbf{\bibinfo{volume}{84}}, \bibinfo{pages}{064604}
  (\bibinfo{year}{2011}).

\bibitem[{\citenamefont{Standy\l{}o et~al.}(2013)\citenamefont{Standy\l{}o,
  Acosta, Angulo, Berjillos, Duenas, Golovkov, Keeley, Keutgen, Martel,
  Mazzocco et~al.}}]{Standy13}
\bibinfo{author}{\bibfnamefont{L.}~\bibnamefont{Standy\l{}o}},
  \bibinfo{author}{\bibfnamefont{L.}~\bibnamefont{Acosta}},
  \bibinfo{author}{\bibfnamefont{C.}~\bibnamefont{Angulo}},
  \bibinfo{author}{\bibfnamefont{R.}~\bibnamefont{Berjillos}},
  \bibinfo{author}{\bibfnamefont{J.~A.} \bibnamefont{Duenas}},
  \bibinfo{author}{\bibfnamefont{M.~S.} \bibnamefont{Golovkov}},
  \bibinfo{author}{\bibfnamefont{N.}~\bibnamefont{Keeley}},
  \bibinfo{author}{\bibfnamefont{T.}~\bibnamefont{Keutgen}},
  \bibinfo{author}{\bibfnamefont{I.}~\bibnamefont{Martel}},
  \bibinfo{author}{\bibfnamefont{M.}~\bibnamefont{Mazzocco}},
  \bibnamefont{et~al.}, \bibinfo{journal}{Phys. Rev. C}
  \textbf{\bibinfo{volume}{87}}, \bibinfo{pages}{064603}
  (\bibinfo{year}{2013}).

\bibitem[{\citenamefont{Kolata}(2002)}]{Kol02}
\bibinfo{author}{\bibfnamefont{J.~J.} \bibnamefont{Kolata}},
  \bibinfo{journal}{Eur. Phys. J. A} \textbf{\bibinfo{volume}{13}},
  \bibinfo{pages}{117} (\bibinfo{year}{2002}).

\bibitem[{\citenamefont{Aguilera et~al.}(2001)\citenamefont{Aguilera, Kolata,
  Becchetti, DeYoung, Hinnefeld, Horv\'ath, Lamm, Lee, Lizcano, Martinez-Quiroz
  et~al.}}]{Aguilera2001}
\bibinfo{author}{\bibfnamefont{E.~F.} \bibnamefont{Aguilera}},
  \bibinfo{author}{\bibfnamefont{J.~J.} \bibnamefont{Kolata}},
  \bibinfo{author}{\bibfnamefont{F.~D.} \bibnamefont{Becchetti}},
  \bibinfo{author}{\bibfnamefont{P.~A.} \bibnamefont{DeYoung}},
  \bibinfo{author}{\bibfnamefont{J.~D.} \bibnamefont{Hinnefeld}},
  \bibinfo{author}{\bibfnamefont{A.}~\bibnamefont{Horv\'ath}},
  \bibinfo{author}{\bibfnamefont{L.~O.} \bibnamefont{Lamm}},
  \bibinfo{author}{\bibfnamefont{H.-Y.} \bibnamefont{Lee}},
  \bibinfo{author}{\bibfnamefont{D.}~\bibnamefont{Lizcano}},
  \bibinfo{author}{\bibfnamefont{E.}~\bibnamefont{Martinez-Quiroz}},
  \bibnamefont{et~al.}, \bibinfo{journal}{Phys. Rev. C}
  \textbf{\bibinfo{volume}{63}}, \bibinfo{pages}{061603(R)}
  (\bibinfo{year}{2001}).

\bibitem[{\citenamefont{Escrig et~al.}(2007)\citenamefont{Escrig,
  S\'anchez-Ben\'{\i}tez, Moro, \'Alvarez, Andr\'es, Angulo, Borge, Cabrera,
  Cherubini, Demaret et~al.}}]{ESCRIG2007}
\bibinfo{author}{\bibfnamefont{D.}~\bibnamefont{Escrig}},
  \bibinfo{author}{\bibfnamefont{A.~M.} \bibnamefont{S\'anchez-Ben\'{\i}tez}},
  \bibinfo{author}{\bibfnamefont{A.~M.} \bibnamefont{Moro}},
  \bibinfo{author}{\bibfnamefont{M.~A.~G.} \bibnamefont{\'Alvarez}},
  \bibinfo{author}{\bibfnamefont{M.~V.} \bibnamefont{Andr\'es}},
  \bibinfo{author}{\bibfnamefont{C.}~\bibnamefont{Angulo}},
  \bibinfo{author}{\bibfnamefont{M.~J.~G.} \bibnamefont{Borge}},
  \bibinfo{author}{\bibfnamefont{J.}~\bibnamefont{Cabrera}},
  \bibinfo{author}{\bibfnamefont{S.}~\bibnamefont{Cherubini}},
  \bibinfo{author}{\bibfnamefont{P.}~\bibnamefont{Demaret}},
  \bibnamefont{et~al.}, \bibinfo{journal}{Nucl. Phys. A}
  \textbf{\bibinfo{volume}{792}}, \bibinfo{pages}{2} (\bibinfo{year}{2007}).

\bibitem[{\citenamefont{Navin et~al.}(2004)\citenamefont{Navin, Tripathi,
  Blumenfeld, Nanal, Simenel, Casandjian, de~France, Raabe, Bazin, Chatterjee
  et~al.}}]{Navi04}
\bibinfo{author}{\bibfnamefont{A.}~\bibnamefont{Navin}},
  \bibinfo{author}{\bibfnamefont{V.}~\bibnamefont{Tripathi}},
  \bibinfo{author}{\bibfnamefont{Y.}~\bibnamefont{Blumenfeld}},
  \bibinfo{author}{\bibfnamefont{V.}~\bibnamefont{Nanal}},
  \bibinfo{author}{\bibfnamefont{C.}~\bibnamefont{Simenel}},
  \bibinfo{author}{\bibfnamefont{J.~M.} \bibnamefont{Casandjian}},
  \bibinfo{author}{\bibfnamefont{G.}~\bibnamefont{de~France}},
  \bibinfo{author}{\bibfnamefont{R.}~\bibnamefont{Raabe}},
  \bibinfo{author}{\bibfnamefont{D.}~\bibnamefont{Bazin}},
  \bibinfo{author}{\bibfnamefont{A.}~\bibnamefont{Chatterjee}},
  \bibnamefont{et~al.}, \bibinfo{journal}{Phys. Rev. C}
  \textbf{\bibinfo{volume}{70}}, \bibinfo{pages}{044601}
  (\bibinfo{year}{2004}).

\bibitem[{\citenamefont{Chatterjee et~al.}(2008)\citenamefont{Chatterjee,
  Navin, Shrivastava, Bhattacharyya, Rejmund, Keeley, Nanal, Nyberg, Pillay,
  Ramachandran et~al.}}]{Chat08}
\bibinfo{author}{\bibfnamefont{A.}~\bibnamefont{Chatterjee}},
  \bibinfo{author}{\bibfnamefont{A.}~\bibnamefont{Navin}},
  \bibinfo{author}{\bibfnamefont{A.}~\bibnamefont{Shrivastava}},
  \bibinfo{author}{\bibfnamefont{S.}~\bibnamefont{Bhattacharyya}},
  \bibinfo{author}{\bibfnamefont{M.}~\bibnamefont{Rejmund}},
  \bibinfo{author}{\bibfnamefont{N.}~\bibnamefont{Keeley}},
  \bibinfo{author}{\bibfnamefont{V.}~\bibnamefont{Nanal}},
  \bibinfo{author}{\bibfnamefont{J.}~\bibnamefont{Nyberg}},
  \bibinfo{author}{\bibfnamefont{R.~G.} \bibnamefont{Pillay}},
  \bibinfo{author}{\bibfnamefont{K.}~\bibnamefont{Ramachandran}},
  \bibnamefont{et~al.}, \bibinfo{journal}{Phys. Rev. Lett.}
  \textbf{\bibinfo{volume}{101}}, \bibinfo{pages}{032701}
  (\bibinfo{year}{2008}).

\bibitem[{\citenamefont{{Yu. E. Penionzhkevich} et~al.}(2007)\citenamefont{{Yu.
  E. Penionzhkevich}, {R. A. Astabatyan}, {N. A. Demekhina}, {G. G. Gulbekian},
  {R. Kalpakchieva}, {A. A. Kulko}, {S. M. Lukyanov}, {E. R. Markaryan}, {V. A.
  Maslov}, {Yu. A. Muzychka} et~al.}}]{Peni07}
\bibinfo{author}{\bibnamefont{{Yu. E. Penionzhkevich}}},
  \bibinfo{author}{\bibnamefont{{R. A. Astabatyan}}},
  \bibinfo{author}{\bibnamefont{{N. A. Demekhina}}},
  \bibinfo{author}{\bibnamefont{{G. G. Gulbekian}}},
  \bibinfo{author}{\bibnamefont{{R. Kalpakchieva}}},
  \bibinfo{author}{\bibnamefont{{A. A. Kulko}}},
  \bibinfo{author}{\bibnamefont{{S. M. Lukyanov}}},
  \bibinfo{author}{\bibnamefont{{E. R. Markaryan}}},
  \bibinfo{author}{\bibnamefont{{V. A. Maslov}}},
  \bibinfo{author}{\bibnamefont{{Yu. A. Muzychka}}}, \bibnamefont{et~al.},
  \bibinfo{journal}{Eur. Phys. J. A} \textbf{\bibinfo{volume}{31}},
  \bibinfo{pages}{185} (\bibinfo{year}{2007}).

\bibitem[{\citenamefont{Lemasson et~al.}(2011)\citenamefont{Lemasson, Navin,
  Rejmund, Keeley, Zelevinsky, Bhattacharyya, Shrivastava, Bazin, Beaumel,
  Blumenfeld et~al.}}]{Lema11}
\bibinfo{author}{\bibfnamefont{A.}~\bibnamefont{Lemasson}},
  \bibinfo{author}{\bibfnamefont{A.}~\bibnamefont{Navin}},
  \bibinfo{author}{\bibfnamefont{M.}~\bibnamefont{Rejmund}},
  \bibinfo{author}{\bibfnamefont{N.}~\bibnamefont{Keeley}},
  \bibinfo{author}{\bibfnamefont{V.}~\bibnamefont{Zelevinsky}},
  \bibinfo{author}{\bibfnamefont{S.}~\bibnamefont{Bhattacharyya}},
  \bibinfo{author}{\bibfnamefont{A.}~\bibnamefont{Shrivastava}},
  \bibinfo{author}{\bibfnamefont{D.}~\bibnamefont{Bazin}},
  \bibinfo{author}{\bibfnamefont{D.}~\bibnamefont{Beaumel}},
  \bibinfo{author}{\bibfnamefont{Y.}~\bibnamefont{Blumenfeld}},
  \bibnamefont{et~al.}, \bibinfo{journal}{Phys. Lett. B}
  \textbf{\bibinfo{volume}{697}}, \bibinfo{pages}{454} (\bibinfo{year}{2011}).

\bibitem[{\citenamefont{Lemasson et~al.}(2009)\citenamefont{Lemasson,
  Shrivastava, Navin, Rejmund, Keeley, Zelevinsky, Bhattacharyya, Chatterjee,
  de~France, Jacquot et~al.}}]{Lema09}
\bibinfo{author}{\bibfnamefont{A.}~\bibnamefont{Lemasson}},
  \bibinfo{author}{\bibfnamefont{A.}~\bibnamefont{Shrivastava}},
  \bibinfo{author}{\bibfnamefont{A.}~\bibnamefont{Navin}},
  \bibinfo{author}{\bibfnamefont{M.}~\bibnamefont{Rejmund}},
  \bibinfo{author}{\bibfnamefont{N.}~\bibnamefont{Keeley}},
  \bibinfo{author}{\bibfnamefont{V.}~\bibnamefont{Zelevinsky}},
  \bibinfo{author}{\bibfnamefont{S.}~\bibnamefont{Bhattacharyya}},
  \bibinfo{author}{\bibfnamefont{A.}~\bibnamefont{Chatterjee}},
  \bibinfo{author}{\bibfnamefont{G.}~\bibnamefont{de~France}},
  \bibinfo{author}{\bibfnamefont{B.}~\bibnamefont{Jacquot}},
  \bibnamefont{et~al.}, \bibinfo{journal}{Phys. Rev. Lett.}
  \textbf{\bibinfo{volume}{103}}, \bibinfo{pages}{232701}
  (\bibinfo{year}{2009}).

\bibitem[{\citenamefont{Marqu\'{\i}nez-Dur\'an
  et~al.}(2018)\citenamefont{Marqu\'{\i}nez-Dur\'an, Martel,
  S\'anchez-Ben\'{\i}tez, Acosta, Aguado, Berjillos, Pinto, Garc\'{\i}a,
  Due\~nas, Rusek et~al.}}]{Marq18}
\bibinfo{author}{\bibfnamefont{G.}~\bibnamefont{Marqu\'{\i}nez-Dur\'an}},
  \bibinfo{author}{\bibfnamefont{I.}~\bibnamefont{Martel}},
  \bibinfo{author}{\bibfnamefont{A.~M.} \bibnamefont{S\'anchez-Ben\'{\i}tez}},
  \bibinfo{author}{\bibfnamefont{L.}~\bibnamefont{Acosta}},
  \bibinfo{author}{\bibfnamefont{J.~L.} \bibnamefont{Aguado}},
  \bibinfo{author}{\bibfnamefont{R.}~\bibnamefont{Berjillos}},
  \bibinfo{author}{\bibfnamefont{A.~R.} \bibnamefont{Pinto}},
  \bibinfo{author}{\bibfnamefont{T.}~\bibnamefont{Garc\'{\i}a}},
  \bibinfo{author}{\bibfnamefont{J.~A.} \bibnamefont{Due\~nas}},
  \bibinfo{author}{\bibfnamefont{K.}~\bibnamefont{Rusek}},
  \bibnamefont{et~al.}, \bibinfo{journal}{Phys. Rev. C}
  \textbf{\bibinfo{volume}{98}}, \bibinfo{pages}{034615}
  (\bibinfo{year}{2018}).

\bibitem[{\citenamefont{Kolata et~al.}(2016)\citenamefont{Kolata, Guimar\~aes,
  and Aguilera}}]{Kola16}
\bibinfo{author}{\bibfnamefont{J.~J.} \bibnamefont{Kolata}},
  \bibinfo{author}{\bibfnamefont{V.}~\bibnamefont{Guimar\~aes}},
  \bibnamefont{and} \bibinfo{author}{\bibfnamefont{E.~F.}
  \bibnamefont{Aguilera}}, \bibinfo{journal}{Eur. Phys. J. A}
  \textbf{\bibinfo{volume}{52}}, \bibinfo{pages}{123} (\bibinfo{year}{2016}).

\bibitem[{\citenamefont{Kolata}(2001)}]{Kol01}
\bibinfo{author}{\bibfnamefont{J.~J.} \bibnamefont{Kolata}},
  \bibinfo{journal}{Phys. Rev. C} \textbf{\bibinfo{volume}{63}},
  \bibinfo{pages}{061604} (\bibinfo{year}{2001}).

\end{thebibliography}


\end{document}